\documentclass[pdflatex, 
sn-nature,
referee]{sn-jnl}

\usepackage{mleftright}
\usepackage{comment}
\usepackage{placeins}

\usepackage{graphicx}%
\usepackage{multirow}%
\usepackage{amsmath,amssymb,amsfonts}%
\usepackage{amsthm}%
\usepackage{mathrsfs}%
\usepackage[title]{appendix}%
\usepackage{xcolor}%
\usepackage{textcomp}%
\usepackage{manyfoot}%
\usepackage{booktabs}%
\usepackage{algorithm}%
\usepackage{algorithmicx}%
\usepackage{algpseudocode}%
\usepackage{listings}%
\usepackage{subfigure}
\usepackage{upgreek}     

\newcommand*{\Poloptwo}[2]{{{}\hat{P}^{\vphantom{\dagger}}_{}}{\vphantom{P}}_{#1}^{#2}}

\newcommand*{\Poldagtwo}[2]{{{}\hat{P}^{\dagger}}{\vphantom{P}}_{#1}^{#2}}

\newcommand*{\Poltwo}[2]{P{\vphantom{P}}_{#1}^{#2}}

\newcommand*{\Nextwo}[2]{N_{#1}^{#2}}

\newcommand*{\Cextwo}[2]{C_{#1}^{#2}}
\newcommand*{\Kextwo}[2]{K_{#1}^{#2}}

\newcommand*{\vdag}[1]{v^{\dagger}_{#1}}

\newcommand*{\cndag}[1]{c_{#1}^{\vphantom{\dagger}}}

\newcommand*{\vdagtwo}[2]{{v^{\dagger}_{}}{\vphantom{v}}_{#1}^{#2}}
\newcommand*{\vndagtwo}[2]{{v_{}^{\vphantom{\dagger}}}{\vphantom{v}}_{#1}^{#2}}
\newcommand*{\cdagtwo}[2]{{c_{}^{\dagger}}{\vphantom{c}}_{#1}^{#2}}
\newcommand*{\cndagtwo}[2]{{c_{}^{\vphantom{\dagger}}}{\vphantom{c}}_{#1}^{#2}}

\newcommand{\ExWFtwo}[2]{\varphi_{#1}^{#2}}
\newcommand{\ExWFstartwo}[2]{\varphi^*{\vphantom{\varphi}}_{#1}^{#2}}

\begin{document}

\title{Dissecting intervalley coupling mechanisms in monolayer transition metal dichalcogenides}


\author*[1]{\fnm{Oleg} \sur{Dogadov}}\email{oleg.dogadov@polimi.it}

\author[2]{\fnm{Henry} \sur{Mittenzwey}}

\author[1]{\fnm{Micol} \sur{Bertolotti}}

\author[3]{\fnm{Nicholas} \sur{Olsen}}

\author[4]{\fnm{Thomas} \sur{Deckert}}

\author[1,5]{\fnm{Chiara} \sur{Trovatello}}

\author[3]{\fnm{Xiaoyang} \sur{Zhu}}

\author[4]{\fnm{Daniele} \sur{Brida}}

\author[1,6]{\fnm{Giulio} \sur{Cerullo}}

\author[2]{\fnm{Andreas} \sur{Knorr}}

\author*[1]{\fnm{Stefano} \sur{Dal Conte}}\email{stefano.dalconte@polimi.it}

\affil[1]{\orgdiv{Department of Physics}, \orgname{Politecnico di Milano}, \orgaddress{\street{Piazza Leonardo da Vinci 32}, \city{Milan}, \postcode{20133}, \country{Italy}}}

\affil[2]{\orgdiv{Institut f\"ur Theoretische Physik, Nichtlineare Optik und Quantenelektronik}, \orgname{Technische Universit\"at Berlin}, \orgaddress{\street{Hardenbergstra{\ss}e 36}, \city{Berlin}, \postcode{10623}, \country{Germany}}}

\affil[3]{\orgdiv{Department of Chemistry}, \orgname{Columbia University}, \orgaddress{\city{New York}, \postcode{10027}, \state{NY}, \country{USA}}}

\affil[4]{\orgdiv{Department of Physics and Materials Science}, \orgname{University of Luxembourg}, \orgaddress{\street{162a Avenue de la Fa\"iencerie}, \city{Luxembourg}, \postcode{1511}, \country{Luxembourg}}}

\affil[5]{\orgdiv{Department of Mechanical Engineering}, \orgname{Columbia University}, \orgaddress{\city{New York}, \postcode{10027}, \state{NY}, \country{USA}}}

\affil[6]{\orgname{CNR-IFN}, \orgaddress{\street{Piazza Leonardo da Vinci 32}, \city{Milan}, \postcode{20133}, \country{Italy}}}


\abstract{
Monolayer (1L) transition metal dichalcogenides (TMDs) provide a unique opportunity to control the valley degree of freedom of optically excited charge carriers due to the spin-valley locking effect. 
Despite extensive studies of the valley-contrasting physics, stimulated by perspective valleytronic applications, a unified picture of competing intervalley coupling processes in 1L-TMDs is lacking.
Here, we apply broadband helicity-resolved transient absorption to explore exciton valley polarization dynamics in 1L-WSe${}_2$. 
By combining experimental results with microscopic simulations, we dissect individual intervalley coupling mechanisms and reveal the crucial role of phonon-assisted scattering in the fast decay of the A exciton circular dichroism and the formation of the dichroism of opposite polarity for the B exciton. 
We further provide a consistent description of the valley depolarization driven by an intervalley-exchange-activating momentum-dark Dexter process and indicate the presence of efficient single electron spin-flip mechanisms. 
Our study brings us closer to a complete understanding of exciton dynamics in TMDs.
}




\maketitle

\section*{Introduction} \label{sec1}

The presence of two energy degenerate valleys with opposite orbital magnetic moments and Berry curvatures in monolayer (1L) group VI transition metal dichalcogenides (TMDs) has stimulated a wide interest in the study of the valley-contrasting physics in these materials \cite{xiao2012, xu2014, schaibley2016, liu2019, zhao2021}. 
The particular interest for this class of compounds is explained by their outstanding optical properties. 
In addition to the presence of the two non-equivalent valleys at $K$ and $K^\prime$ points of the Brillouin zone, the reduced dimensionality of 1L-TMDs, their direct gap, and a large spin-orbit interaction give rise to tightly bound exciton states \cite{chernikov2014}, which can be selectively addressed in $K$ and $K^\prime$ valleys with light of opposite helicity \cite{sallen2012, zeng2012, mak2018}.
However, for a successful use of the valley degree of freedom in prospective devices \cite{schaibley2016, liu2019}, a way to generate and control a robust valley polarization, i.e., population difference of the two valleys, must be achieved.
An understanding of the microscopic processes governing the intra- and intervalley dynamics in 1L-TMDs after optical excitation becomes therefore a critical factor.  

Ultrafast optical spectroscopies provide powerful methods to study exciton dynamics in 1L-TMDs \cite{dalconte2020}.
Transient Faraday or Kerr rotation spectroscopy \cite{zhu2014, plechinger2017} and helicity-resolved pump-probe (also referred to as transient circular dichroism (CD)) spectroscopy \cite{mai2014, mai2014cd, wang2018} are the two widely used techniques that allow one to address the dynamics of valley degrees of freedom. 
These studies have revealed an ultrafast coupling of the valleys in 1L-TMDs \cite{lloyd2021}, which has been attributed to different many-body correlations \cite{mai2014, schmidt2016, berghauser2018}.
Intervalley electron-hole exchange interaction based on Coulomb-interaction-induced energy transfer has been initially proposed to explain the fast loss of valley polarization \cite{selig2019ultrafast, yu2014valley, combescot2023ab}. 
However, phonon-assisted intervalley electron scattering has also been suggested to be an efficient process in 1L-TMDs \cite{kioseoglou2016, plechinger2017, molina-sanchez2017, carvalho2017, brem2020, he2020, bae2022, selig2018dark}.
Moreover, a Dexter-like intervalley interaction based on Coulomb-interaction-induced charge transfer that mixes the A and B excitonic transitions from opposite valleys might also contribute to the valley depolarization process \cite{berghauser2018, bernal2018exciton}, whereas a coupling of A and B excitons in the same valley can be induced by intravalley exchange \cite{guo2019exchange}.
The spin-valley physics becomes particularly complicated for the case of spin-dark materials, such as WSe${}_2$ or WS${}_2$, where the role of different mechanisms is still debated \cite{huang2017temporal, koutensky2023ultrafast, raiber2022ultrafast, molina-sanchez2017}.
Although many of the above mentioned interaction mechanisms have been studied before on their own, it is still unclear how their combined action affects the valley (de-)polarization dynamics in 1L-TMDs and whether there is a predominant mechanism.

In this work, we apply broadband femtosecond transient CD spectroscopy to study the spin-valley relaxation dynamics in 1L-$\mathrm{WSe_2}$. 
By simultaneously characterizing the A and B excitons, we are able to resolve complex dynamics of the two transitions, and unveil the presence of multiple inter- and intravalley relaxation channels. 
We simulate the valley-resolved exciton dynamics by solving the microscopic Heisenberg equations of motion for coherent transitions and incoherent occupations within a unit operator method~\cite{ivanov1993, katsch2018}. 
We find that initial decay of the A exciton CD signal and the formation of the B exciton CD of the opposite sign are mainly due to a fast intervalley phonon-assisted scattering, while the valley depolarization on a longer timescale is governed by an exchange-activating momentum-dark Dexter interaction, which has not yet been described in the literature. 
A strong temperature dependence of the CD dynamics suggests an unequal equilibration of electrons and holes, due to an additional electron single spin-flip mechanism. 
Thanks to the excellent agreement of numerical simulations and experimental results, our combined study is able to evaluate the role of multiple excitonic interaction processes governing the spin-valley dynamics in 1L-TMDs.


\section*{Results}

\subsection*{Experiment}

\begin{figure}[ht]
    \centering
    \includegraphics[width=.42\textwidth]{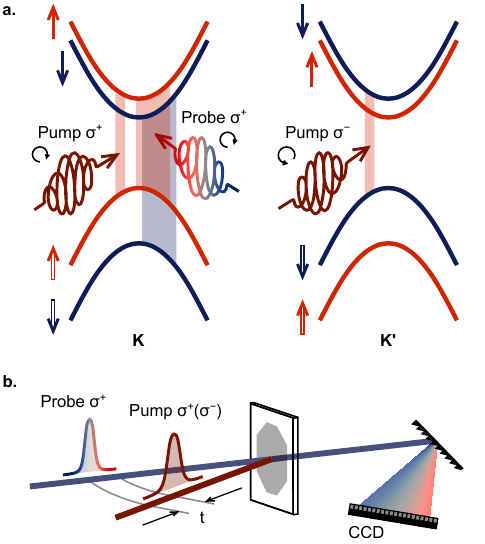}
    \caption{
    \textbf{Transient circular dichroism.} 
    \textbf{a} Valence and conduction band spin splitting in $K$ and $K^\prime$ points of the Brillouin zone of 1L-$\mathrm{WSe_2}$, giving rise to A and B excitons. 
    Color coding shows bands with the same spin orientation, indicated with solid arrows for electrons and double arrows for holes.
    Right ($\sigma^+$) and left ($\sigma^-$) circularly polarized pulses couple to two nonequivalent valleys of a 1L-TMD.
    Broadband $\sigma^+$ probe covers both A and B excitons at the $K$ point.
    \textbf{b} Diagram of a transient CD experiment.  
    }
    \label{fig:cartoon}
\end{figure}

Experiments are performed on a large-area 1L-$\mathrm{WSe_2}$ on $\mathrm{SiO_2}$ substrate, fabricated with gold-assisted mechanical exfoliation technique (see Methods section for details). 
We use femtosecond differential transmission (DT) spectroscopy with circularly polarized pulses to access intra- and intervalley dynamics. 
We resonantly excite the A exciton $1s$ state with either right ($\sigma^+$) or left ($\sigma^-$) circularly polarized pulses to selectively generate A excitons in either $K$ or $K^\prime$ valley.
The $\sim\! 25$~meV energy width of the pump pulses limits the temporal resolution of our experiments to $\sim \! 100$~fs, as confirmed by measuring the instrumental response function (IRF).  
A broadband $\sigma^+$-polarized supercontinuum probe tracks the dynamics of the A and B excitons in the $K$ valley. 
Figure~\ref{fig:cartoon} illustrates the principle and implementation of the experiment.
The CD is calculated as the difference of the two DT signals following excitation with opposite helicities:
\begin{equation}
    \mathrm{CD} = \left. \frac{\Delta T}{T}\right|_{\sigma^+\sigma^+} \! -
                    \left. \frac{\Delta T}{T}\right|_{\sigma^-\sigma^+}
    \label{eq:cd}
\end{equation}
Here, $\Delta T / T := (T_\mathrm{pump \ on} - T_\mathrm{pump \ off}) / T_\mathrm{pump \ off}$, where $T_\mathrm{pump \ on}$ and $T_\mathrm{pump \ off}$ are transmission signals with and without pump, respectively. 
The double indices in Equation~(\ref{eq:cd}) specify the pump (first) and the probe (second) polarizations. 

%


\subsection*{Transient circular dichroism}

\begin{figure}[ht]
    \centering
    \includegraphics[width=1.0\textwidth]{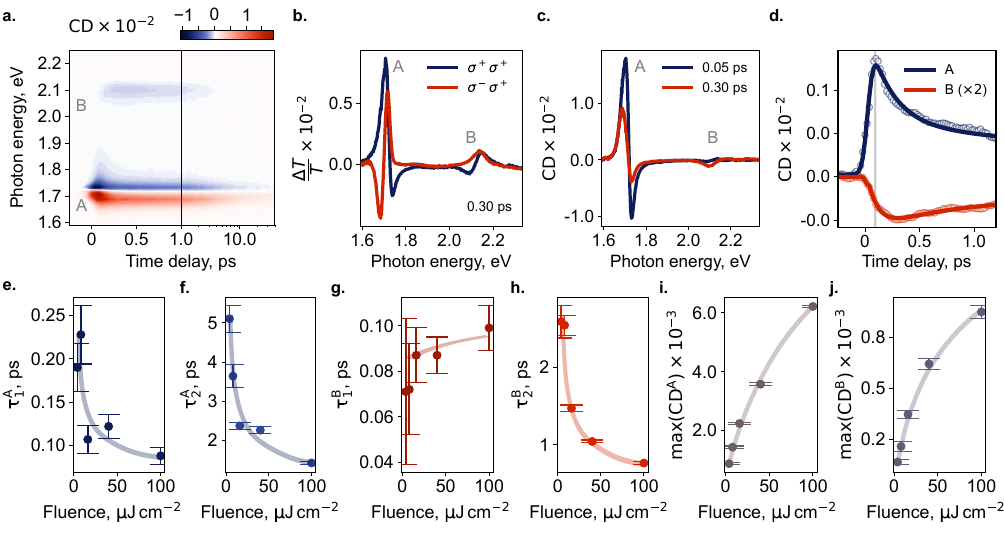}
    \caption{
    \textbf{Valley-resolved dynamics.} 
    \textbf{a} Pseudo-color energy-time CD pump-probe map, as defined in Equation~(\ref{eq:cd}), 8~$\upmu\mathrm{J \, cm^{-2}}$ pump fluence.
    \textbf{b} Transient spectra at 0.3~ps for two pump circular polarizations.
    \textbf{c} Transient spectra of the CD map (\textbf{a}) at select time delays.
    \textbf{d} Spectrally averaged temporal CD traces of A and B excitons. 
    Vertical line shows the time delay of maximum A exciton CD magnitude.
    Dots correspond to experimental points, solid lines show the fit. 
    \textbf{e}--\textbf{j} Fluence dependence. 
    \textbf{e},\textbf{f} Fluence dependence of the two lifetime components for A exciton. 
    \textbf{g},\textbf{h} Same for B exciton. 
    \textbf{i},\textbf{j} Fluence dependence of magnitudes of A and B CD signals, respectively. 
    Solid lines in panels \textbf{e}--\textbf{j} are guides to eye.
    All measurements are performed at 77~K.
    }
    \label{fig:exp}
\end{figure}

Figure~\ref{fig:exp}a reports transient CD data at 77 K, as a function of probe photon energy $E$ and pump-probe delay $\Delta t$.
The CD signal is calculated from the two DT maps measured for opposite pump helicities, as defined in Equation~(\ref{eq:cd}).
In this measurement, the sample is excited in resonance with the A exciton with 8~$\upmu\mathrm{J \, cm^{-2}}$ fluence, which corresponds to $\sim \! 8 \times 10^{11}$~cm$^{-2}$ photocarrier density.
The map reveals the presence of complex transient signals around A and B exciton resonances, i.e., 1.7~eV and 2.1~eV, respectively. 
The A exciton CD signal shows a dispersive line shape, changing its sign from positive to negative, as the probe photon energy increases. 
The B exciton signal, instead, is dominated by a negative component.
The origin of the CD signal is illustrated in Figure~\ref{fig:exp}b,c.
Panel~(\textbf{b}) depicts DT spectra at 0.30~ps time delay for two pump helicities ($\sigma^+$/$\sigma^-$) resonant with the A exciton and broadband $\sigma^+$ probe.
Both A and B exciton resonances have different line shapes and signal strengths, giving rise to distinct CD signals. 
Transient CD spectra in Figure~\ref{fig:exp}c, obtained from the map (\textbf{a}) at two time delays, show that the rapid decrease of the A exciton CD signal within first 0.30~ps upon photoexcitation is accompanied by a delayed formation of a B exciton CD signal. 

The transient CD signal thus has a complex shape, originating from valley population imbalance and many-body effects \cite{sie2015, schmidt2016, deckert2025coherent_tmp}. 
To isolate the temporal evolution of the degree of valley polarization we perform a spectral integration of the CD signal over 300~meV energy ranges around the exciton peaks. 
Spectrally averaged A and B exciton transient CD traces are shown in Figure~\ref{fig:exp}d, which isolates the Pauli-blocking-induced bleaching dynamics.
The graph reveals opposite signs of the A and B excitons' CD signals, as previously observed in 1L-$\mathrm{WS_2}$ \cite{berghauser2018, wang2018}.
Furthermore, whereas the signal of A exciton is formed within the pump pulse duration, the B exciton CD has a finite formation time, as shown with a vertical line, marking the time delay at which the trace of the A exciton reaches its maximum value. 
The traces are fitted with a biexponential function, convoluted with a single Gaussian profile $\mathcal{G}(\sigma = 60 \ \mathrm{fs})$, which represents the IRF of the system:
\begin{equation}
    \mathrm{CD} \approx H \cdot \left[ \sum_{i=1}^{2} A_i \, \exp \left( -\frac{\Delta t}{\tau_i} \right) + C \right] * \mathcal{G}(60 \, \mathrm{fs})
    \label{eq:decay}
\end{equation}
Here, $H$ is a Heaviside step function, $C$ is a constant offset. 
We find that the fast relaxation component of the A exciton signal $\tau_1^\mathrm{A} = 100 \pm 10$~fs and the formation signal of the B exciton CD $\tau_1^\mathrm{B} = 90 \pm 20$~fs are of the same order, suggesting that common processes govern their dynamics. 
The slower relaxation times of both signals are found to be on a picosecond timescale with a faster decay of the B exciton signal. 
Figure~\ref{fig:exp}e--j reports a pump fluence dependence of the first (``short'') and the second (``long'') time constants for the CD signals of the two excitons, as well as the maximum values of the CD. 
The magnitudes of the signals show a sub-linear scaling with pump fluence, which is related to the saturation of the DT signals for both pump helicities \cite{kumar2014, seo2016}. 
As the pump fluence increases, the depolarization dynamics, characterized by the second time constants $\tau_2$, extracted for the CD signals of both excitons becomes faster, as shown in Figure~\ref{fig:exp}f,h. 
Such behavior was previously observed in different 1L-TMDs \cite{dalconte2015, mahmood2018} and it was explained by an enhanced exciton-exciton exchange interaction at increased carrier densities \cite{mahmood2018}.  
We comment on the possible origin of this effect in the Discussion section.
Importantly, for all investigated fluences the decay time of the B exciton CD signal is shorter than that of the A exciton. 
We also find that as the pump fluence increases, the first time constant $\tau_1$ becomes shorter for the A exciton, whereas for the B exciton it remains unchanged within the precision of our measurements.
The higher uncertainty in the definition of $\tau_1$ is due to the fact that its value is on the same order as the duration of the pump pulse.
We note that even for the highest applied pump fluence the photocarrier density is approximately an order of magnitude lower than the expected Mott transition threshold \cite{chernikov2015population, dendzik2020observation}.

To confirm the main observations reported above, we perform two-color time-resolved Faraday rotation measurements.
In these experiments, the dynamics of the valley population imbalance is measured by the photoinduced rotation of the linearly-polarized probe polarization, originating from the different refractive indexes experienced by its $\sigma^+$ and $\sigma^-$ components \cite{dalconte2015, hsu2015}. 
The performed measurements support our observations, demonstrating the opposite sign of the valley polarization for A and B excitons, as well as the shorter depolarization time for the B exciton.
The outline of the experimental setup and the experimental results are presented in Supplementary Text, Sections~I.A.2 an III.A, respectively.


\subsection*{Temperature dependence}

\begin{figure}[ht]
    \centering
    \includegraphics[width=1.0\textwidth]{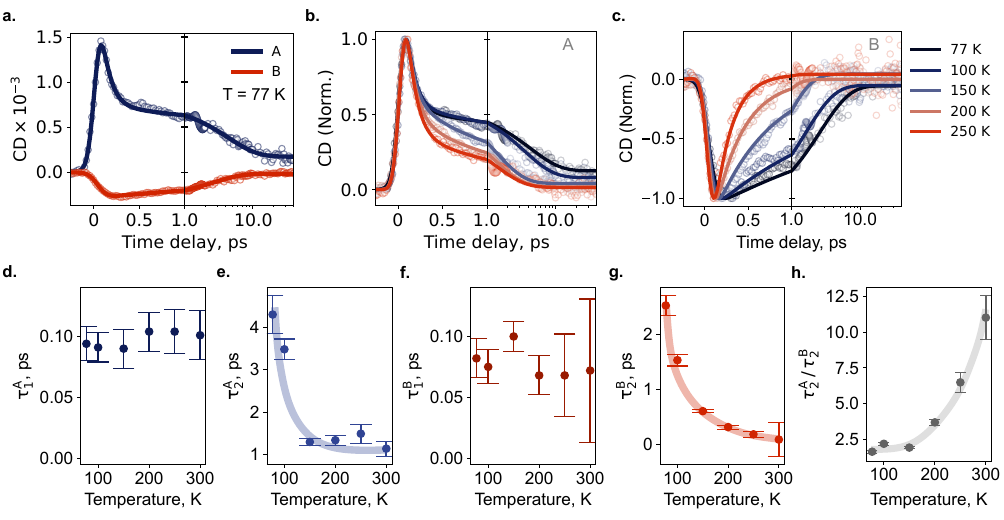}
    \caption{
    \textbf{Temperature dependence.} 
    \textbf{a} Spectrally averaged CD signals of A and B excitons at 77~K. 
    \textbf{b}, \textbf{c} Temperature dependence of A and B exciton CD signals, respectively. 
    In panels \textbf{a}--\textbf{c}, dots are experimental points, and solid lines show a biexponential fit. 
    \textbf{d}--\textbf{h} Temperature dependence of the time constants $\tau_1$ (\textbf{d}) and $\tau_2$ (\textbf{e}) for CD signal of A exciton, $\tau_1$ (\textbf{f}) and $\tau_2$ (\textbf{g}) for CD signal of B exciton, and the ratio of the second lifetime components $\tau_2^\mathrm{A} / \tau_2^\mathrm{B}$ for two excitons (\textbf{f}).
    Solid lines in panels \textbf{e}, \textbf{g}, and \textbf{h} are guides to eye. 
    }
    \label{fig:tdep}
\end{figure}

We further explore the temperature dependence of the CD signal, following resonant excitation of the A exciton with 8~$\upmu\mathrm{J \, cm^{-2}}$ pump fluence.
Figure~\ref{fig:tdep}a compares CD traces of A and B excitons at 77~K.
The temperature dependencies of the CD signals of A and B excitons are separately presented in panels (\textbf{b}) and (\textbf{c}), respectively. 
While the fast depolarization component of the A exciton CD and the formation time of the B exciton CD are weakly affected by temperature, the long decay time for both A and B excitons becomes progressively shorter as the temperature increases.
The results of the lifetime analysis of the CD signals approximated with the function in Equation~(\ref{eq:decay}) are summarized in Figure~\ref{fig:tdep}d--h.

The increase of the depolarization rate for A exciton at higher temperatures shown in Figure~\ref{fig:tdep}e is in agreement with the previously reported studies for 1L-TMDs \cite{dalconte2015, zhu2014, mahmood2018}, in which it has been explained by the Coulomb exchange interaction \cite{mahmood2018} or phonon-mediated intervalley spin-flip transitions \cite{molina-sanchez2017}.
In our measurements, a similar effect is also observed for the B exciton CD (Figure~\ref{fig:tdep}g). 
However, the temperature effect on the depolarization dynamics of the B exciton is more prominent.
Figure~\ref{fig:tdep}h shows how the ratio of the depolarization times for A and B excitons $\tau_2^\mathrm{A} / \tau_2^\mathrm{B}$ changes with temperature. 
Within the accuracy of the lifetime analysis, the ratio remains unchanged up to ca.~150~K. 
As the temperature increases further, the depolarization rate of the B exciton CD accelerates faster than that of the A exciton, and at 300~K, its depolarization time is an order of magnitude shorter. 
Such anomalous temperature dependence of the B exciton polarization suggests that, apart from the above mentioned processes, an additional temperature dependent spin-flip mechanism must be considered to correctly describe this effect. 
Note that, while the lifetime of the B exciton CD signal experiences such strong temperature dependence, the $\Delta T / T$ signal itself does not show a significant temperature dependence of the relaxation time (see Supplementary Text, Section~III.B).


\subsection*{Theoretical model}

The experimentally observed opposite sign of the CD signals for A and B excitons suggests the presence of strong intervalley coupling mechanisms.
As mentioned before, coherent interactions related to the pump-induced excitonic polarization are expected to play an important role in the spin-valley dynamics of 1L-TMDs.   
However, the fact that the B exciton CD signal shows a finite formation time implies the presence of efficient incoherent mechanisms, which involve excitonic populations in the two valleys.
For a comprehensive description of the spin-valley dynamics it is crucial therefore to consider the combined action of different interaction mechanisms. 

To obtain an in-depth explanation of the experimentally observed CD signatures and to disentangle the role of phonon-assisted intervalley scattering and Coulomb intervalley coupling processes, we perform microscopic simulations for the exciton dynamics.
The calculation scheme for the temporal evolution of the coherent excitonic polarization $P$ (see Equations~\eqref{eq:P} and~\eqref{eq:P_pump} in the Methods), and the incoherent exciton occupation $N$ (see Equations~\eqref{eq:N} and~\eqref{eq:N_pump} in the Methods), is based on a Heisenberg equations of motion approach in a correlation expansion of electron-hole pairs (see Supplementary Text, Section~I.B.1) \cite{ivanov1993, katsch2018, fricke1996transport}.

\begin{figure}[ht]
    \centering
    \includegraphics[width=.42\textwidth]{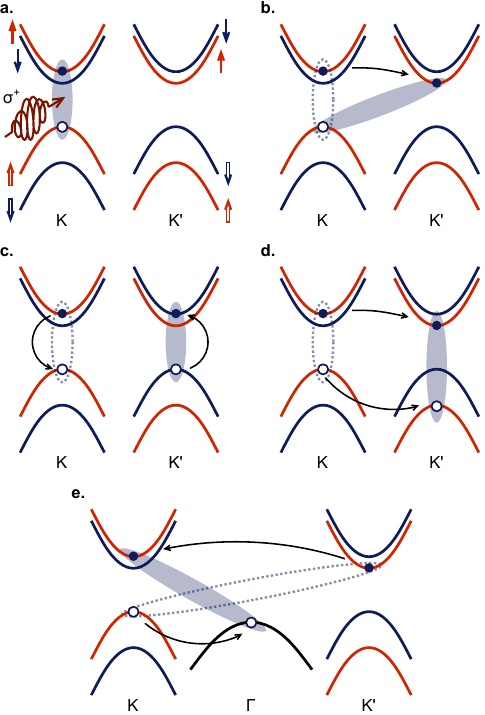}
    \caption{
    \textbf{Select intervalley exciton occupation scattering processes in 1L-$\mathbf{WSe_2}$ in $\mathbf{K}$ and $\mathbf{K^\prime}$ valleys.} 
    We show the processes that follow $\sigma^+$ excitation, creating A excitons in $K$ valley, as shown in panel \textbf{a}.
    Solid and double colored arrows indicate electron and hole spin states, respectively.
    \textbf{b} Intervalley phonon-assisted electron scattering. 
    \textbf{c} Resonant intervalley exchange.
    \textbf{d}--\textbf{e} Incoherent Dexter interaction on intravalley (\textbf{d}) and intervalley (\textbf{e}) occupations.
    Dashed ellipses illustrate initial exciton states, filled ellipses show final exciton states.
    }
    \label{fig:scattering_digarams}
\end{figure}

Before delving into the theoretical model in detail, we first establish the optical observable to create a one-to-one correspondence to the measured signals.
Since the aim of this work is to examine the physical mechanisms which determine the dynamics of the excitonic bleaching (Pauli-blocking) signal (see for instance Figure~\ref{fig:exp}d) we disregard all coherent pump-induced spectral signatures due to Coulomb interaction such as excitation-induced energy renormalization and broadening \cite{katsch2020exciton, trovatello2022disentangling, katsch2019theory, deckert2025coherent_tmp} or excitation-induced broadening due to exciton-phonon interaction \cite{katzer2023exciton}. 
Upon energy-averaging the experimentally measured DT spectra around the respective excitonic resonances, the influence of these excitation-induced spectral signatures on the probed signal vanishes and only Pauli-blocking due to pump-induced coherent and incoherent excitonic occupations remains.
Therefore, the energy-averaged DT signal $\Delta T(E^i, \Delta t)$ around excitonic energy $E^i$ for $i=\text{A},\text{B}$ dependent on the pump-probe delay $\Delta t$ by using Equation~\eqref{eq:P_mic} can be expressed as \cite{policht2023time}:
\begin{align}
    \Delta T(E^i, \Delta t) \sim\left|\mathbf d^{vc,i}\right|^2\sum_{i^{\prime},\mathbf Q}D_{\mathbf Q}^{i,i^{\prime}}\left(|P^{i^{\prime}}(\Delta t)|^2\delta_{\mathbf Q,\mathbf 0} + N_{\mathbf Q}^{i^{\prime}}(\Delta t)\right).
    \label{eq:DeltaTSignal}
\end{align}
In Equation~\eqref{eq:DeltaTSignal}, $i$ denotes the probed transition and $i^{\prime}$ (A and B exciton) denotes the respective coherent ($|P^{i^{\prime}}(\Delta t)|^2$) at $\mathbf Q=\mathbf 0$ or incoherent ($N_{\mathbf Q}^{i^{\prime}}$) excitonic occupation with respect to the center-of-mass momentum $\mathbf Q$ contributing to the Pauli-blocking mediated by the bleaching form factors $D_{\mathbf Q}^{i,i^{\prime}}$. 
These factors are given in Equation~(\ref{eq:DOverlaps}) of the Methods and consistently take into account the fermionic substructure of the exciton via encoding Pauli-blocking by the individual holes and electrons. 
Therefore, the DT signal depends on the oscillator strength of the probed excitonic transition $i$, as well as on the occurrence of electron and/or hole Pauli-blocking due to excitonic occupations in the state $i^{\prime}$, as soon as the electron and/or the hole of the pump-induced excitonic occupation overlap with any electron and/or hole of the probed transition in the state $i$.
This is illustrated exemplarily in Figure~\ref{fig:scattering_digarams}a,b: After the optical pump in Figure~\ref{fig:scattering_digarams}a, bleaching of the B transition at $K^{\prime}$ occurs.
From Equation~\eqref{eq:DeltaTSignal}, the simulated CD signal is obtained using Equation~\eqref{eq:cd}.
In the following, we discuss the CD dynamics in terms of the bleaching-inducing momentum-integrated occupations $N$. 
However, in our simulations we always take into account the full momentum-dependent overlap between the bleaching form factors $D_{\mathbf Q}$ (defined in Equation~\eqref{eq:DOverlaps} in Methods) and the excitonic occupations $N_{\mathbf Q}$.

Having established the relation of the coherent exciton polarization $P$ and occupations $N_{\mathbf Q}$ to the measured spectra, we turn to the microscopic model. 
Sketches in Figure~\ref{fig:scattering_digarams} illustrate the possible exciton intervalley scattering processes after $\sigma^+$ excitation of the A exciton at the $K$ point (Figure~\ref{fig:scattering_digarams}a), which might occur in the following and primarily contribute to the dynamics of the CD signals of A and B excitons.
In our calculations, we take into account phonon-assisted incoherent exciton formation and thermalization, as well as intervalley scattering \cite{selig2018dark,thranhardt2000quantum} (panel (\textbf{b})), Coulomb-mediated intervalley exchange scattering (F\"orster coupling, double spin flips) \cite{selig2019ultrafast, yu2014valley, maialle1993exciton} (panel (\textbf{c})),  and Coulomb-mediated direct intervalley scattering of intra- and intervalley excitonic occupations (Dexter coupling) \cite{berghauser2018, bernal2018exciton} (panels (\textbf{d}) and (\textbf{e}), respectively).
We also examine the role of the coherent Dexter-like intervalley exciton coupling, which instantaneously mixes A and B excitons in opposite valleys \cite{berghauser2018}. 
Details to all matrix elements and equations of motion can be found in the Supplementary Text, Sections~I.B.2--5.

\begin{figure}[ht]
    \centering
    \includegraphics[width=.42\textwidth]{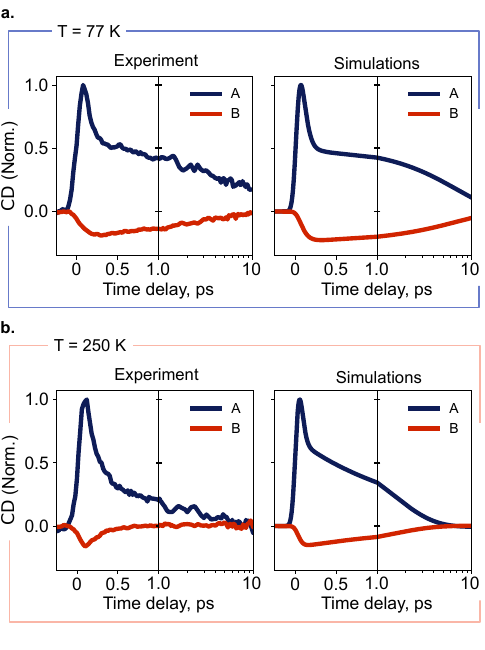}
    \caption{
    \textbf{Experimental and simulated results.}
    Comparison of normalized A and B exciton CD traces for 77~K (\textbf{a}) and 250~K (\textbf{b}).
    }
    \label{fig:th_exp}
\end{figure}

Figure~\ref{fig:th_exp} compares the measured and simulated CD signals at 77~K (\textbf{a}) and 250~K (\textbf{b}). 
For both the temperatures, a good agreement between the experiment and the theory is obtained.  
The simulations accurately capture the initial fast decay of the A exciton signal and the delayed formation of the B exciton CD, as well as the relative strengths of dichroism signals of A and B excitons. 
Moreover, the overall decay of the CD signal, which is experimentally shown to be faster at higher temperatures, is correctly reproduced in the simulations. 
Our calculations, however, do not reproduce the faster decay of the B exciton CD signal compared to that of the A exciton.

Figure~\ref{fig:theory_components} shows in detail the modeled CD dynamics along with individual contributions of selected coupling mechanisms.
In the next section, we present an in-depth discussion of the role of different physical processes and provide an explanation to the main experimental observables.


\section*{Discussion}

Upon pumping the A exciton transition in the $K$ valley shown in Figure~\ref{fig:scattering_digarams}a and a subsequent phonon-assisted dephasing of coherent excitonic polarizations into incoherent excitonic intravalley occupations, spin-preserving phonon-assisted intervalley scattering leads to the formation of energetically lower intervalley occupations \mbox{$K^{}_\Uparrow$-$K^{\prime}_\uparrow$} occupations (in this notation, the first (second) index denotes the valley spin of the hole $\Uparrow,\Downarrow$ (electron $\uparrow,\downarrow$) of the corresponding exciton), as shown in Figure~\ref{fig:scattering_digarams}b.
This process occurs mainly via the spontaneous emission of phonons, which explains the relative insensitivity of the fast decay of the A exciton CD to temperature (see Figure~\ref{fig:tdep}b). 
In the simulations, we also include scattering to intermediate states at the \mbox{$\Lambda$/$\Lambda^{\prime}$} valleys (not shown in Figure~\ref{fig:scattering_digarams}), which causes an overall speed-up of the phonon-assisted intervalley scattering. 
Overall, the CD signal of the probed A transition is positive, since intervalley \mbox{$K^{}_\Uparrow$-$K^{\prime}_\uparrow$} occupations cause a bleaching in the $\sigma^+\sigma^+$ pump-probe configuration at the $K$ valley due to the hole remaining there, whereas intervalley \mbox{$K^{\prime}_\Downarrow$-$K^{}_\downarrow$} occupations induce no bleaching in the $\sigma^-\sigma^+$ pump-probe configuration, because the hole is located at the $K^{\prime}$ valley and the electron at the $K$ valley has an opposite spin compared to the probed transition, see Equation~(\ref{eq:cd}).
The probed B transition, instead, experiences a bleaching in the $\sigma^-\sigma^+$ configuration due to the electron part of the intervalley \mbox{$K^{\prime}_\Downarrow$-$K^{}_\downarrow$} excitonic occupation, whereas the bleaching of the probed B signal in the $\sigma^+\sigma^+$ configuration is zero, since \mbox{$K^{}_\Uparrow$-$K^{\prime}_\uparrow$} occupations do not contribute to the Pauli-blocking in the probed B transition at the $K$ valley (see Figure~\ref{fig:scattering_digarams}b). 
This explains the experimentally observed negative CD signal, see Equation~(\ref{eq:cd}).

\begin{figure}[ht]
    \centering
    \includegraphics[width=.42\textwidth]{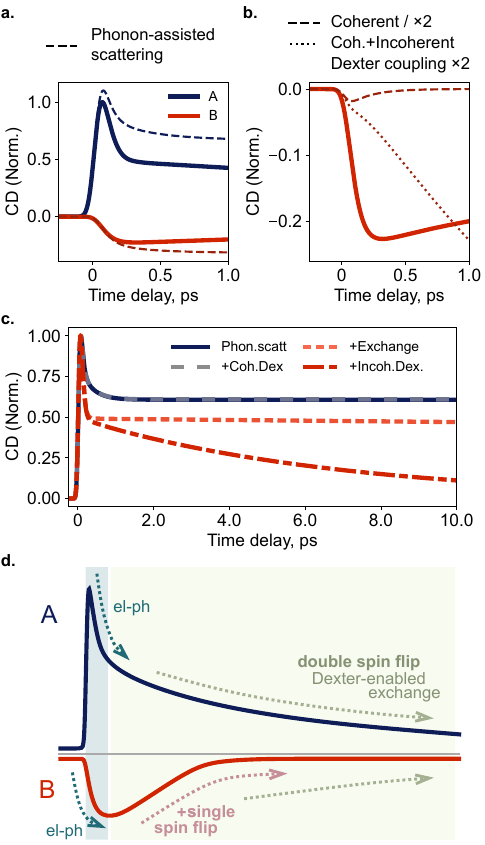}
    \caption{
    \textbf{Contribution of select processes to the CD signal at 77~K.}
    \textbf{a} Comparison of total simulated CD traces with the phonon-assisted intervalley scattering contribution to the CD. 
    \textbf{b} Total simulated B exciton CD compared to the effect of Dexter coupling (coherent and incoherent) alone.
    The Dexter coupling traces are magnified by a factor of 2 for clarity.
    In both panels, the traces are normalized to the maximum A exciton CD value (for complete simulations). 
    \textbf{c} Contributions of phonon-assisted scattering, coherent Dexter coupling, intervalley exchange, and incoherent intervalley Dexter coupling to the overall decay of the A exciton CD signal.
    \textbf{d} Summary of the main processes.
    A cartoon schematically shows the CD dynamics of A and B excitons in 1L-WSe${}_2$ and highlights the key processes, governing the dynamics.
    See details in text.
    }
    \label{fig:theory_components}
\end{figure}

In Figure~\ref{fig:theory_components}a, the formation of the A and B CD signals is depicted for the full simulations (solid lines) and when only phonon-assisted scattering processes are taken into account (dashed lines). 
Since these traces are qualitatively similar for both A and B excitons, we conclude that phonon-assisted spin-preserving intervalley scattering of electrons is the most relevant mechanism of the initial decay of the A exciton CD signal and the delayed formation of the B exciton CD of opposite polarity (see also the results of calculations at 250~K in Supplementary Figure~19).

Importantly, according to our calculations, coherent Dexter coupling \cite{berghauser2018} cannot explain the observed behavior:
Although due to this mechanism the B excitonic transition in the $K^\prime$ valley is instantaneously excited when the A exciton at the $K$ valley is resonantly optically pumped, this process is by more than an order of magnitude too inefficient to generate the CD signal for the B exciton in Figure~\ref{fig:theory_components}b.
Due to the large energetic mismatch of A and B excitons, the off-resonantly Dexter-excited B excitonic polarization just follows the optically excited A exciton polarization adiabatically and almost no  dephasing of coherent B excitonic polarization into incoherent B excitonic population occurs.
We also find that the incoherent Dexter coupling (illustrated in Figure~\ref{fig:scattering_digarams}d), which involves excitonic populations instead of polarizations, causes a delayed build-up of the B signal, which is stronger than that obtained by considering the coherent one only.
However, the combination of coherent and incoherent Dexter coupling results in a CD signal which is too small and with a too long rise time to explain the measured CD dynamics, as shown in Figure~\ref{fig:theory_components}b.

Now, we discuss the overall decay of the CD signals, i.e.,\ the valley depolarization dynamics. 
In contrast to the initial fast decay of the A signal and the delayed formation of the B signal, the overall decay of both CD signals cannot be explained by phonon-assisted scattering alone, and additional spin flip processes need to be taken into account.
Figure~\ref{fig:theory_components}c depicts the impact of different exciton scattering processes on the valley depolarization dynamics of the A CD signal. 
Since only double-spin-flip processes are included in our simulations (the total spin of electron and hole is conserved during all the considered scattering mechanisms), the effect of the discussed processes on the B exciton CD decay, within our calculations, is identical.
We note that, due to the above mentioned large energetic mismatch between A and B excitons, the coherent Dexter interaction does not give any sizable loss of the A exciton CD signal, as seen in Figure~\ref{fig:theory_components}c.

Although the intervalley exchange is a double spin-flip process, as shown in Figure~\ref{fig:scattering_digarams}c, it does not lead to a valley depolarization on its own due to its suppression when momentum-dark intervalley occupations (shown in Figure~\ref{fig:scattering_digarams}b) dominate \cite{selig2020}, in contrast to the case of spin-bright materials, such as 1L-MoSe${}_2$ \cite{selig2019ultrafast}. 
It becomes therefore necessary to destabilize the energetically favorable intervalley occupation.

According to our calculations, the most significant impact on the valley depolarization is provided by the additional incoherent momentum-dark Dexter interaction on intervalley occupations, shown in Figure~\ref{fig:scattering_digarams}e, which is a different mechanism compared to the Dexter interaction between intravalley polarizations and occupations \cite{berghauser2018, bernal2018exciton, jasinski2024control}.
Although it does not cause a valley depolarization on its own, since it conserves the spin, 
momentum-dark intervalley Dexter coupling (Figure~\ref{fig:scattering_digarams}e) leads to an activation of double-spin-flip-inducing intervalley exchange (Figure~\ref{fig:scattering_digarams}c) by breaking the quasi-stable configuration of intervalley $K_{\Uparrow}^{}$-$K^{\prime}_{\uparrow}$-excitons: A fraction of the $K_{\Uparrow}^{}$-$K^{\prime}_{\uparrow}$-excitons is constantly upconverted into $\Gamma_{\Uparrow}^{}$-$K_{\uparrow}^{}$-excitons in a first step, which then feed into $K_{\Uparrow}^{}$-$K_{\uparrow}^{}$-excitons via phonon-assisted hole scattering, where intervalley exchange subsequently acts upon.
This complex interplay then leads to an overall temperature-dependent decay of both CD signals, which excellently reproduces the measured decay of the A signal.
A detailed explanation of the depolarization mechanism can be found in the Supplementary Text, Section~II.
To the best of our knowledge, the momentum-dark Dexter interaction has not yet been explored in literature, despite its major importance.
We also note that since the valley depolarization results from the complex interplay of several processes, we expect that at higher fluences additional contributions, which are quadratic in the excitonic occupations $\propto \! N^2$ or higher, will contribute to all processes considered, which would explain the observed acceleration of the depolarization dynamics, reported in Figure~\ref{fig:exp}e,f.
However, to the best of our knowledge, no density-dependent excitonic theory, which combines phonon-assisted scattering \cite{katzer2023exciton} and Coulomb intra- and interband scattering, exists so far.

We note that our model does not capture the experimentally observed faster decay of the B exciton CD signal compared to that of the A exciton. 
This difference in the CD decay dynamics implies that spin-down (spin-up) electrons at the $K$ ($K^\prime$) valley, which govern the B exciton CD signal, equilibrate faster than the corresponding holes, which are monitored by the A exciton CD signal. 
This effect can possibly be due to the Rashba spin-orbit interaction, which induces single spin flips, as soon as an out-of-plane electric field is present in a sample \cite{book:Spin_orbit_coupling_Winkler2003,mittenzwey2025rashba}.
Due to the much larger valence band splitting compared to the conduction band splitting, spin hybridization of electrons is more likely compared to spin hybridization of holes. 
Thus, phonon-assisted spin flips between electrons are expected to be much faster than phonon-assisted scattering between holes \cite{wang2018}.
Although several mechanisms that can potentially generate an out-of-plane electric field in 1L-TMDs have been proposed (including image charges at dielectric boundaries \cite{mittenzwey2025many,slobodeniuk2016spin}, local strain \cite{zhuang2019intrinsic}, non-thermal \cite{schlipf2021dynamic} or chiral \cite{lagarde2024efficient,chan2025exciton,zhang2022ab} phonons), the evaluation of the applicability of these mechanisms to our system of interest is beyond the scope of the present work. 

Figure~\ref{fig:theory_components}d summarizes the discussion. 
A and B exciton signals show opposite valley polarization. 
The fast dynamics (light blue area), i.e., the rapid decay of the A exciton CD, as well as the delayed formation of the B exciton CD, is mainly defined by the efficient intervalley electron-phonon scattering (Figure~\ref{fig:scattering_digarams}b). 
The long (depolarization) dynamics (light green area) results from a complex interplay of different processes. 
The overall loss of valley polarization is due to the Coulomb-mediated intervalley exchange, which is a double spin flip process (Figure~\ref{fig:scattering_digarams}c).
This process is enabled by the momentum-dark incoherent Dexter interaction acting on intervalley occupations (Figure~\ref{fig:scattering_digarams}e).
We further rationalize that additional phonon-assisted spin flips between electrons contribute to the faster decay of the B exciton CD signal.

In our work, we have focused on spin-valley dynamics in 1L-WSe${}_2$.
It is important to note, however, that the conclusions provided here are relevant also to other 1L-TMDs, and the developed theoretical approach can be applied to other systems.
We argue that it is necessary to consider the discussed processes for a correct description of exciton dynamics in other spin-dark materials, such as 1L-WS${}_2$.
For the case of 1L-MoS${}_2$, we expect the single spin flip for electrons to be even more efficient than in the studied case, due to the crossing of the spin-split conduction bands close to the $K$ and $K^\prime$ points of the Brillouin zone \cite{kormanyos2015k, liu2013threeband}. 
Our results are also consistent with the current understanding of the valley dynamics in 1L-MoSe${}_2$, which is a spin-bright material \cite{feierabend2021}: Since phonon-assisted intervalley electron scattering is energetically unfavorable, no intervalley exciton population is created, and therefore a very fast valley depolarization is observed within ca.~1~ps \cite{raiber2022ultrafast, jeong2020valley} due to the intervalley exchange \cite{selig2019ultrafast}, which remains fully active through the course of time.

In conclusion, we have performed a combined experimental-theoretical study of intervalley coupling processes in 1L-$\mathrm{WSe_2}$ and assessed their role in the spin-valley relaxation processes. 
By applying broadband transient CD spectroscopy we have revealed notable dichroism signals for both A and B excitons, implying the presence of strong intervalley coupling mechanisms. 
The opposite sign of the CD signals for A and B excitons, the initial fast decay of the A exciton CD signal and the delayed formation of the B exciton CD signal can be explained by intervalley phonon-assisted electron scattering.
To correctly describe the decay of the CD signals, i.e., the overall valley depolarization dynamics, instead, one needs to consider additional effects:
While intervalley exchange alone is not sufficient to explain the measured valley depolarization dynamics, the complex interplay of the exchange-activating momentum-dark Dexter interaction with phonon-assisted scattering provides a very good agreement of the simulations with the experimentally observed results. 
Additionally, we have found experimentally that the B exciton CD signal shows a much faster decay, compared to the A exciton one, indicating that the electron spin-valley imbalance gets equilibrated faster than that of holes. 
Prospective studies comprise a detailed investigation of possible mechanisms for single-spin processes in 1L-TMDs.


\section*{Methods} \label{sec11}

\bmhead{Experimental set-up} 
Time-resolved measurements are performed in transmission geometry (see Supplementary Figure~1).
The same setup with different detection schemes is used for the transient CD and for the TRFR measurements.
The setup is seeded by a regeneratively amplified Ti-sapphire laser (Coherent, Libra), which provides 100-fs pulses at 1.55~eV at a 2~kHz repetition rate.
Narrowband pump pulses are generated in a home-built noncollinear optical parametric amplifier (NOPA), tuned to the energy of the A exciton 1s state (1.70~eV), and modulated by a mechanical chopper at half repetition rate of the amplifier (1~kHz).
The broadband supercontinuum probe is generated by focusing 1.55~eV beam in a 2~mm-thick $\mathrm{Al_2 O_3}$ plate.
The spectral region below 1.59~eV is eliminated by a short-pass filter.
The polarization states of pump and probe beams are controlled separately by a combination of linear polarizers and achromatic waveplates.  
The two beams are focused on a sample inside a cryostat (Oxford Instruments) at almost normal incidence with a small angle ($\sim \! 5^\circ$) between them. 
The transmitted circularly polarized probe is detected by a spectrometer with a silicon CCD.  
The time traces are acquired by scanning relative delays between pump and probe pulses by a motorized delay stage. 


\bmhead{Sample preparation} 
Macroscopic 1L-$\mathrm{WSe_2}$ is exfoliated via the gold tape exfoliation method \cite{liu2020}. 
Bulk $\mathrm{WSe_2}$, grown via chemical vapor transport (CVT), have been purchased from HQ Graphene. 
Gold tape is prepared by evaporating gold onto a polished silicon wafer before spin coating a protective layer of polyvinylpyrrolidone (PVP). 
A 150~nm layer of gold is deposited onto the polished silicon wafer at a rate of 0.05~nm/s (Angstrom Engineering EvoVac Multi-Process thin film deposition system).
A solution of PVP (40,000mw Alpha Aesar), ethanol, and acetonitrile with a 2:9:9 mass ratio is spun onto the gold surface of the wafer (1000~rpm, 1000~rpm/s acceleration, 2~min) before curing on a hot plate (150~$^\circ$C, 5~min). 
Using thermal release tape (Semiconductor Equipment Corp. Revalpha RA-95LS(N)), 
the gold tape is removed from the silicon wafer and is pressed onto the surface of a bulk TMD crystal to exfoliate a large area monolayer. 
After placing the 1L-$\mathrm{WSe_2}$ onto the final substrate, the sample assembly is heated on a hot plate at 130~$^\circ$C to remove the thermal release tape. 
The sample assembly is soaked in deionized water for 3 hours and in acetone for 1 hour to remove any remaining polymer residue. 
Then, the gold is then etched for 5~minutes in a KI/$\mathrm{I_2}$ gold etchant solution (Iodine, 99.99\%, Alfa Aesar; potassium iodide, 99.0\%, Sigma-Aldrich, and deionized water with a 4:1:40 mass ratio). 
The sample is soaked in deionized water for 2 hours, rinsed in isopropanol, and dried with $\mathrm{N_2}$.


\bmhead{Theoretical calculations}
The differential transmission signal is obtained by developing the excitonic Bloch equations for the excitonic transitions:
\begin{equation}
    \Poltwo{\mu}{\xi,s} = \left\langle \Poloptwo{\mu,\mathbf Q=\mathbf 0}{\xi,\xi,s,s}\right\rangle_c,
    \label{eq:P}
\end{equation}
by taking into account the light-matter interaction
within a correlation expansion in electron-hole pairs in the excitonic picture \cite{katsch2018theory,fricke1996transport}
up to the third order in the optical field $\mathbf E$. The excitonic transition operator is defined by:
\begin{equation}
    \Poloptwo{\mu,\mathbf Q}{\xi,\xi^{\prime},s,s^{\prime}} = \sum_{\mathbf q}\ExWFstartwo{\mu,\mathbf q}{\xi,\xi^{\prime},s,s^{\prime}}\vdagtwo{\mathbf q-\beta\mathbf Q}{\xi,s}\cndagtwo{\mathbf q+\alpha\mathbf Q}{\xi^{\prime},s^{\prime}},
\end{equation}
where $\mu$ is the excitonic quantum number, $\mathbf Q$ is the center-of-mass momentum, the index pairs $\xi,\xi^{\prime}$ and $s,s^{\prime}$ denote the valley and spin of the hole (first index) and electron (second index) forming the corresponding exciton.
We denote the spins of the hole as $s=\Uparrow,\Downarrow$ and the spins of the electron as $s^{\prime}=\uparrow,\downarrow$. Moreover, $\mathbf q$ is the relative momentum, $\ExWFtwo{\mu,\mathbf q}{\xi,\xi^{\prime},s,s^{\prime}}$ is the excitonic wave function solving the Wannier equation, $\alpha$, $\beta$ are the effective-mass ratios of the corresponding exciton and $\vdag{}$ ($\cndag{}$) is the valence band electron creation (conduction band electron annihilation) operator.
The equations of motion for the probe-induced excitonic transitions $\Poltwo{\mu}{\text{pr},\xi,\xi,s,s}$ read:
\begin{equation}
\begin{split}
	& \mathrm i\hbar\partial_t \Poltwo{\mu}{\text{pr},\xi,\xi,s,s}=\left(E_{\mu}^{\xi,\xi,s,s}-\mathrm i\hbar\gamma_{\mu}^{\xi,\xi,s,s}\right)\Poltwo{\mu}{{\text{pr},}\xi,\xi,s,s} - \mathbf E_{}^{{\text{pr}}}\cdot\mathbf d^{cv,\xi,s}_{}\sum_{\mathbf q}\ExWFstartwo{\mu,\mathbf q}{\xi,\xi,s,s}\\
    &\,+\mathbf E_{}^{{\text{pr}}}\cdot\mathbf d^{cv,\xi,s}_{}
 \sum_{\substack{\nu,\mathbf Q,\xi^{\prime},\xi^{\prime\prime},\\s^{\prime},s^{\prime\prime}}}
 D^{\xi,\xi,s,s;\xi^{\prime},\xi^{\prime\prime},s^{\prime},s^{\prime\prime}}_{\mu,\nu,\mathbf Q}
 \left(\left|\Poltwo{\nu}{\xi^{\prime},\xi^{\prime},s^{\prime},s^{\prime}}\right|^2\delta_{\mathbf Q,\mathbf 0}\delta_{\xi^{\prime},\xi^{\prime\prime}}^{s^{\prime},s^{\prime\prime}}+\Nextwo{\nu,\mathbf Q}{\xi^{\prime},\xi^{\prime\prime},s^{\prime},s^{\prime\prime}}\right),
 \end{split}
 \label{eq:P_mic}
\end{equation}
where $E_{\mu}^{\xi,\xi,s,s}$ is the excitonic energy, $\gamma_{\mu}^{\xi,\xi,s,s}$ is the excitonic non-radiative dephasing, $\mathbf E^{{\text{pr}}}_{} = \mathbf E_0^{{\text{pr}}} - \dfrac{1}{2\epsilon_0n_{\text{ref}}c_0}\partial_t\mathbf P^{{\text{pr}}}_{}$ is the total optical probe field with incoming field $\mathbf E_0^{{\text{pr}}}{ = \mathbf E^{\text{t}}_0 \, \delta(t-\Delta t)}$ (delta-shaped probe pulse with pump-probe delay time $\Delta t$) and macroscopic polarization $\mathbf P^{{\text{pr}}}$ and $\mathbf d^{cv,\xi,s}_{}$ is the interband transition dipole moment \cite{mkrtchian2019theory}. $D^{\xi,\xi,s,s;\xi^{\prime},\xi^{\prime\prime},s^{\prime},s^{\prime\prime}}_{\mu,\nu,\mathbf Q}$ are the bleaching form factors, given by:
\begin{align}
    D^{\xi,\xi,s,s;\xi^{\prime},\xi^{\prime\prime},s^{\prime},s^{\prime\prime}}_{\mu,\nu,\mathbf Q} = \sum_{\mathbf q}\ExWFstartwo{\mu,\mathbf q}{\xi,\xi,s,s}\left(\left|\ExWFtwo{\nu,\mathbf q+\beta\mathbf Q}{\xi^{\prime},\xi^{\prime\prime},s^{\prime},s^{\prime\prime}}\right|^2\delta_{\xi^{\prime},\xi}^{s^{\prime},s}
    +
    \left|\ExWFtwo{\nu,\mathbf q-\alpha\mathbf Q}{\xi^{\prime},\xi^{\prime\prime},s^{\prime},s^{\prime\prime}}\right|^2\delta_{\xi^{\prime\prime},\xi}^{s^{\prime\prime},s}\right),
    \label{eq:DOverlaps}
\end{align}
where the first contribution is related to Pauli-blocking by holes and the second is related to Pauli-blocking by electrons of the corresponding coherently excited pump-induced exciton occupation $\left|\Poltwo{\mu}{\xi^{\prime},\xi^{\prime},s^{\prime},s^{\prime}}\right|^2$ and pump-induced incoherent excitonic occupation $\Nextwo{\nu,\mathbf Q}{\xi^{\prime},\xi^{\prime\prime},s^{\prime},s^{\prime\prime}}$:
\begin{align}
    \Nextwo{\nu,\mathbf Q}{\xi^{\prime},\xi^{\prime\prime},s^{\prime},s^{\prime\prime}} = \left\langle\Poldagtwo{\nu,\mathbf Q}{\xi^{\prime},\xi^{\prime\prime},s^{\prime},s^{\prime\prime}}\Poloptwo{\nu,\mathbf Q}{\xi^{\prime},\xi^{\prime\prime},s^{\prime},s^{\prime\prime}} \right\rangle_c.
    \label{eq:N}
\end{align}
The equations of motion for the pump-induced excitonic transitions read:
\begin{align}
    {\mathrm i\hbar\partial_t \Poltwo{\mu}{\xi,s} = \mathrm i\hbar\partial_t \Poltwo{\mu}{\xi,s}\Big|_0 + \mathrm i\hbar\partial_t \Poltwo{\mu}{\xi,s}\Big|_{\text{opt}} + \mathrm i\hbar\partial_t \Poltwo{\mu}{\xi,s}\Big|_{\text{X-phon}} + \mathrm i\hbar\partial_t \Poltwo{\mu}{\xi,s}\Big|_{\text{Coul,Dex}}, }
    \label{eq:P_pump}
\end{align}
and the equations of motion for the pump-induced excitonic occupations read:
\begin{align}
    {\partial_t\Nextwo{\mu,\mathbf Q}{\xi,\xi^{\prime},s} = \partial_t\Nextwo{\mu,\mathbf Q}{\xi,\xi^{\prime},s}\Big|_{\text{X-phon}} + \partial_t\Nextwo{\mu,\mathbf Q}{\xi,\xi^{\prime},s}\Big|_{\text{Coul,X}} + \partial_t\Nextwo{\mu,\mathbf Q}{\xi,\xi^{\prime},s}\Big|_{\text{Coul,Dex}}. }
    \label{eq:N_pump}
\end{align}
The individual free ``0'' and optical ``op'' contributions are given in Section~I.B.2, the exciton-phonon contributions ``X-phon'' are given in Section~I.B.3, the Coulomb exchange contributions ``Coul,X'' are given in Section~I.B.4, and the Coulomb Dexter contributions ``Coul,Dex'' are given in Section~I.B.5 in the Supplementary Text.
From the probe-induced transition in Eq.~\eqref{eq:P_mic}, the macroscopic probe-induced polarization $\mathbf P^{{\text{pr}}}_{} = \frac{1}{\mathcal A}\sum_{\mu,\xi,s,\mathbf q}\mathbf d^{vc,\xi,s}_{}\ExWFtwo{\mu,\mathbf q}{\xi,\xi,s,s}\Poltwo{\mu}{{\text{pr}},\xi,\xi,s,s}$ can be established, which enables the calculation of the transmission by self-consistently solving Maxwell's equations \cite{knorr1996theory,jahnke1997linear}. The differential transmission $\Delta T$ is then obtained by subtracting the transmission of the probe pulse without the pump pulse from the transmission of the probe pulse with the pump pulse, yielding Equation~\ref{eq:DeltaTSignal} in the main part of the manuscript.


\backmatter

\section*{Supplementary information}

Supplementary Text and Figures 1–13.

\section*{Data availability}

The data supporting the findings of this study are available upon request.

\section*{Acknowledgements}

N.O. acknowledges the assistance of Anthony Calderon and Alex Smirnov in sample preparation.
O.D. and G.C. acknowledge financial support from European Union's NextGenerationEU Investment 1.1, PRIN 2022 PNRR HAPPY [ID P20224AWLB, CUP D53D23016720001].
S.D.C and G.C. acknowledge financial support by the European Union's NextGenerationEU Programme with the I-PHOQS Infrastructure [IR0000016, ID D2B8D520, CUP B53C22001750006] “Integrated infrastructure initiative in Photonic and Quantum Sciences.” 
S.D.C. acknowledges support from the European Union's NextGenerationEU – Investment 1.1, M4C2 - Project n. 2022LA3TJ8 – CUP D53D23002280006.
Fabrication of macroscopic monolayer samples was supported as part of Programmable Quantum Materials, an Energy Frontier Research Center funded by the U.S.~Department of Energy (DOE), Office of Science, Basic Energy Sciences (BES), under award DE-SC0019443.
H.M.~and A.K.~acknowledge financial support from the Deutsche Forschungsgemeinschaft (DFG) through Project~KN~427/11-2,~Project~No.~420760124. A.K.~acknowledges financial support from the DFG through Project~KN~427/15-1,~Project~No.~556436549.
D.B. and T.D. acknowledge support from the FEDER Project LuxUltrafast 2 2023-01-04.
C.T. acknowledges the European Union’s Horizon Europe research and innovation programme under the Marie Skłodowska-Curie PIONEER HORIZON-MSCA-2021-PF-GF grant agreement~No~101066108.

\section*{Author contribution}

O.D.~and S.D.C.~devised the experiment; O.D.~and M.B.~conducted the experiment; O.D., M.B.,~and T.D. analyzed the data; H.M.~performed the theoretical work; N.O.~and C.T.~fabricated and characterized the sample; O.D.~and H.M.~wrote the manuscript with input from all coauthors; X.Y.Z., D.B., A.K., G.C., and S.D.C.~oversaw the project.

\section*{Competing interests}

The authors declare no competing interests.


\newpage

{\Large \textbf{Supplementary Material for: ``Dissecting intervalley coupling mechanisms in monolayer transition metal dichalcogenides''}}

\section{Extended methods}

\subsection{Experimental apparatus}

\subsubsection{Transient circular dichroism}

A diagram of the pump-probe confocal microscope used for transient reflectivity measurements is depicted in Supplementary Figure~\ref{fig:clark_short}.
A detailed description of the setup is provided in the Methods section of the main text.
\begin{figure}[ht]
    \centering
    \includegraphics[width = .7\textwidth]{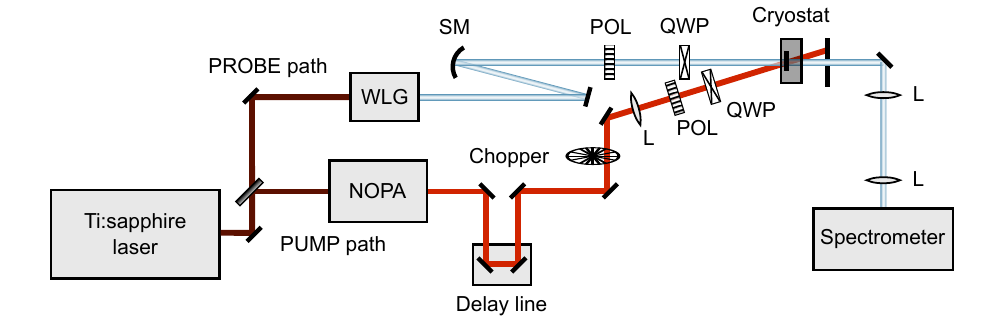}
    \caption{
    \textbf{Transient CD experimental setup.}
    WLG is white light generation unit, SM is a spherical mirror, POL are polarizers, QWP are quarter-wave plates, L are lenses.
    }
    \label{fig:clark_short}
\end{figure}
The white light generation (WLG) units consists of a lens, focusing the fundamental of the laser on a 2~mm-thick $\mathrm{Al_2 O_3}$ plate on a manual translation stage, an iris and a neutral density filter used to adjust the shape and intensity of the focused beam, a spherical mirror used for the collimation of the generated supercontinuum, and a short-pass filter, removing the residual fundamental.
The visible NOPA is designed as explained in Ref.~\cite{cerullo2003}.


\medskip

\subsubsection{Time-resolved Faraday rotation}

For time-resolved Faraday rotation (TRFR) measurements, we use the same setup, as for the CD experiments, with a modified probe path and detection scheme. 
A scheme of the setup is shown in Supplementary Figure~\ref{fig:clark_short_trfr}.
\begin{figure}[ht]
    \centering
    \includegraphics[width = .7\textwidth]{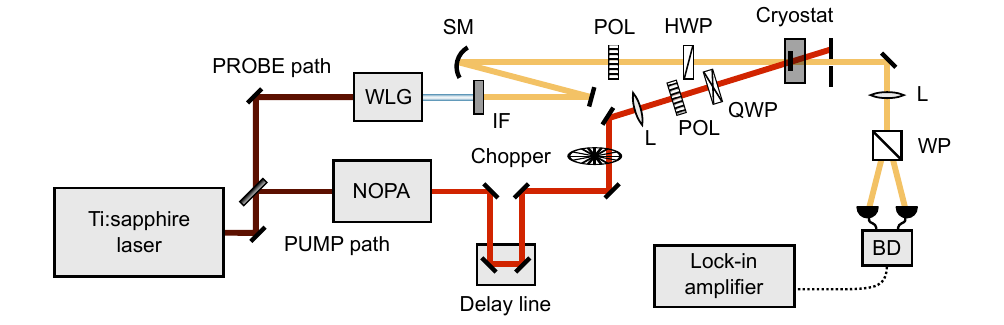}
    \caption{
    \textbf{TRFR experimental setup.}
    WLG is white light generation unit, IF is an interference filter, SM is a spherical mirror, POL are polarizers, QWP is a quarter-wave plate, HWP is a half-wave plate, L are lenses, WP is a Wollaston prism, BD is a balanced detector.
    }
    \label{fig:clark_short_trfr}
\end{figure}
After the WLG, a narrowband probe is selected by an interference filter. 
The rotation of the linearly polarized probe induced by an interaction with a sample is measured in a balanced detection system composed of a Wollaston prism and a balanced silicon photodiode. 
The induced signal imbalance is read by a lock-in amplifier. 

Since a narrowband probe pulse is used in the TRFR measurements, instead of the broadband supercontinuum, the temporal resolution of this method is lower and is limited to ca.~150~fs \cite{polli2010}.


\subsection{Theoretical model}

\subsubsection{Excitonic picture}

\begin{figure}[ht]
    \centering
    \includegraphics[width=0.85\linewidth]{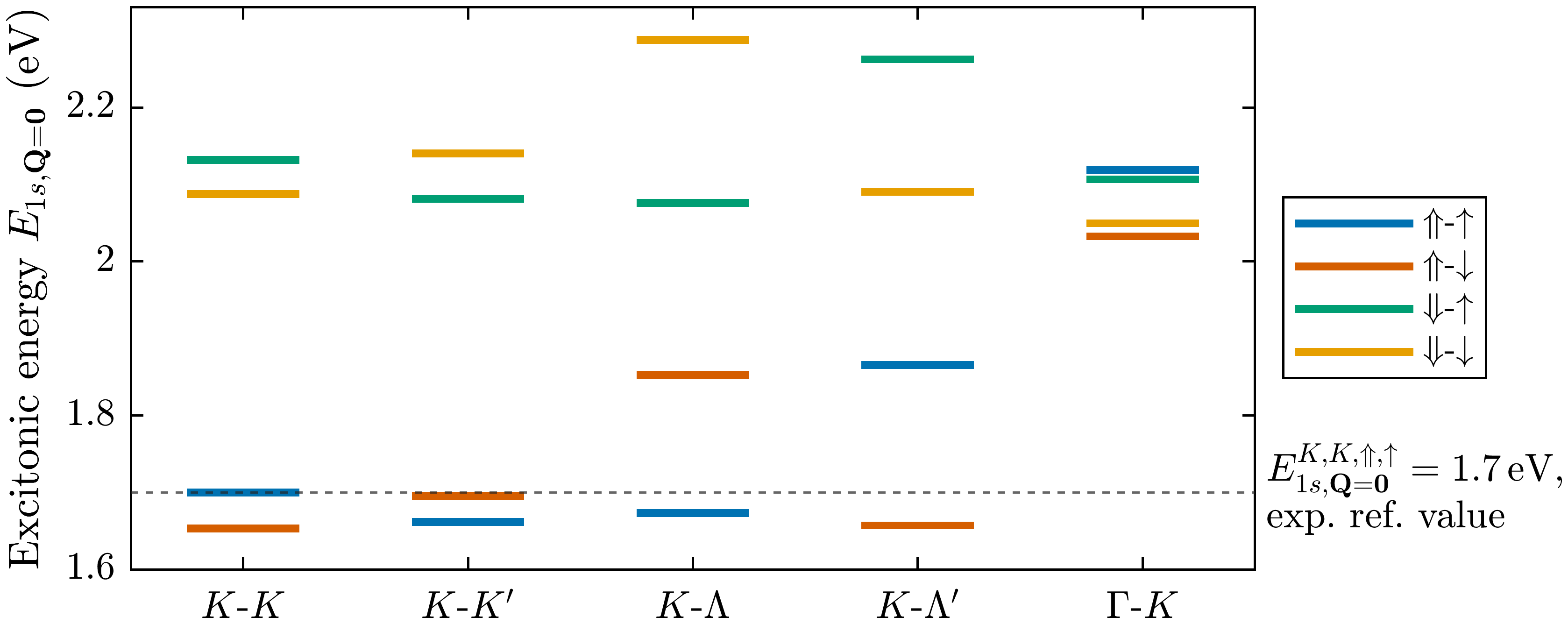}
    \caption{Excitonic energy landscape for WSe$_2$ on SiO$_2$. Note, that the energies of \mbox{$K^{\prime}$-$K$}, \mbox{$K^{\prime}$-$\Lambda$} and \mbox{$\Gamma$-$K^{\prime}$} excitons can be obtained by switching the corresponding valley and spin simultaneously.}
    \label{fig:ExcitonicEnergies}
\end{figure}

We formulate the theory for the pump-induced dynamics in the excitonic picture \cite{katsch2018theory} within a correlation expansion \cite{fricke1996transport} in electron-hole pairs. 
This way, the effect of the electron-hole Coulomb interaction in lowest order is always taken care of.
The excitonic energies $E_{\mu}^{\xi,\xi^{\prime},s,s^{\prime}}$ are obtained by the Wannier equation:
\begin{align}
    \left(E_{\text{gap}}^{\xi,\xi^{\prime},s,s^{\prime}} + \frac{\hbar^2\mathbf q^2}{2}\left(\frac{1}{m_{e}^{\xi^{\prime},s^{\prime}}} + \frac{1}{m_h^{\xi,s}}\right)\right)\ExWFtwo{\mu,\mathbf q}{\xi,\xi^{\prime},s,s^{\prime}} - \sum_{\mathbf q^{\prime}}V_{\mathbf q-\mathbf q^{\prime}}\ExWFtwo{\mu,\mathbf q^{\prime}}{\xi,\xi^{\prime},s,s^{\prime}} = E_{\mu}^{\xi,\xi^{\prime},s,s^{\prime}}\ExWFtwo{\mu,\mathbf q}{\xi,\xi^{\prime},s,s^{\prime}},
    \label{eq:WannierEquation}
\end{align}
where $E_{\text{gap}}^{\xi,\xi^{\prime},s,s^{\prime}}$ is the free-particle band gap, $\mathbf q$ is the relative momentum, $m_{h/e}^{\xi,s}$ are the effective masses of the hole/electron taken from Ref.~\cite{kormanyos2015k} and $\ExWFtwo{\mu,\mathbf q}{\xi,\xi^{\prime},s,s^{\prime}}$ is the excitonic wave function. $V_{\mathbf q}$ is the screened, quantum-confined Coulomb potential given in Eq.~\eqref{eq:V_Screened}. 
The total excitonic dispersion reads:
\begin{align}
    E_{\mu,\mathbf Q}^{\xi,\xi^{\prime},s,s^{\prime}} = E_{\mu}^{\xi,\xi^{\prime},s,s^{\prime}} + \frac{\hbar^2\mathbf Q^2}{2\left(m_h^{\xi,s}+m_{e}^{\xi^{\prime},s^{\prime}}\right)},
\end{align}
where $\mathbf Q$ is the center-of-mass momentum. 
In Fig.~\ref{fig:ExcitonicEnergies}, we depict the energies at zero center-of-mass momentum $E_{\mu,\mathbf Q=\mathbf 0}^{\xi,\xi^{\prime},s,s^{\prime}}$ of all considered exciton configurations.

We account for the excitonic transitions:
\begin{align}
    \Poltwo{\mu}{\xi,s} = \left\langle \Poloptwo{\mu,\mathbf Q=\mathbf 0}{\xi,\xi,s,s}\right\rangle_c,
        \label{eq:P_Def}
\end{align}
the excitonic intravalley ($\xi=\xi^{\prime}$) and intervalley ($\xi\neq\xi^{\prime}$) occupations:
\begin{align}
    \Nextwo{\mu,\mathbf Q}{\xi,\xi^{\prime},s} = \left\langle \Poldagtwo{\mu,\mathbf Q}{\xi,\xi^{\prime},s,s}\Poloptwo{\mu,\mathbf Q}{\xi,\xi^{\prime},s,s}\right\rangle_c,
        \label{eq:N_Def}
\end{align}
the correlations of two intravalley excitons:
\begin{align}
    \Cextwo{\mu,\nu,\mathbf Q}{\xi,\bar\xi,s,s^{\prime}} = \left\langle \Poldagtwo{\mu,\mathbf Q}{\xi,\xi,s,s}\Poloptwo{\nu,\mathbf Q}{\bar{\xi},\bar{\xi},s^{\prime},s^{\prime}}\right\rangle_c,
    \label{eq:C_Def}
\end{align}
and the correlations of two intervalley excitons:
\begin{align}
    \Kextwo{\mu,\nu,\mathbf Q}{\xi,\bar \xi,\Gamma,\xi,s,s^{\prime}} = \left\langle \Poldagtwo{\mu,\mathbf Q}{\xi,\bar{\xi},s,s}\Poloptwo{\nu,\mathbf Q}{\Gamma,\xi,s^{\prime},s^{\prime}}\right\rangle_c,
        \label{eq:K_Def}
\end{align}
in an excitonic Heisenberg equations of motion approach by using bosonic excitonic commutator relations in lowest order \cite{katsch2018theory}:
\begin{align}
    \left[\Poloptwo{\mu,\mathbf Q}{\xi,\xi^{\prime},s,s^{\prime}},\Poldagtwo{\nu,\mathbf K}{\xi^{\prime\prime},\xi^{\prime\prime\prime},s^{\prime\prime},s^{\prime\prime\prime}}\right] = \delta_{\mu,\nu}\delta_{\mathbf Q,\mathbf K}\delta_{\xi,\xi^{\prime\prime}}\delta_{\xi^{\prime},\xi^{\prime\prime\prime}}
    \delta_{s,s^{\prime\prime}}\delta_{s^{\prime},s^{\prime\prime\prime}}.
\end{align}
In the following, we lay out all interaction mechanisms, which we take into account. Since we are only interested in modelling pump-probe bleaching dynamics, we neglect any effects on the spectral characteristics of the pump-induced transition such as Coulomb- or phonon-mediated excitation-induced energy renormalization or excitation-induced dephasing \cite{katsch2020exciton, schmidt2016, katzer2023exciton, deckert2025coherent_tmp}.


\subsubsection{Free excitonic contribution and optical interaction}

The free excitonic Hamiltonian reads:
\begin{align}
    \hat H_0 = \sum_{\mu,\mathbf Q,\xi,\xi^{\prime},s,s^{\prime}}E_{\mu,\mathbf Q}^{\xi,\xi^{\prime},s,s^{\prime}}\Poldagtwo{\mu,\mathbf Q}{\xi,\xi^{\prime},s,s^{\prime}}\Poloptwo{\mu,\mathbf Q}{\xi,\xi^{\prime},s,s^{\prime}},
\end{align}
and the excitonic optical interaction Hamiltonian reads:
\begin{align}
    \hat H_{\text{opt}} = -\sum_{\mu,\mathbf Q,\xi,s}\left( \mathbf E_{\mathbf Q}\cdot \mathbf d^{cv,\xi,s}\sum_{\mathbf q}\ExWFstartwo{\mu,\mathbf q}{\xi,\xi,s,s}\Poldagtwo{\mu,\mathbf Q}{\xi,\xi,s,s} + \mathbf E_{-\mathbf Q}\cdot \mathbf d^{vc,\xi,s}\sum_{\mathbf q}\ExWFtwo{\mu,\mathbf q}{\xi,\xi,s,s}\Poloptwo{\mu,\mathbf Q}{\xi,\xi,s,s}\right),
\end{align}
where $\mathbf E_{\mathbf Q} = \frac{1}{\mathcal A}\int\mathrm dz\,|\xi(z)|^2\mathbf E_{\mathbf Q}(z)$ is the quantum-confined optical field, where $\mathbf E_{\mathbf Q}(z)$ solves the wave equation. Here, $\mathbf d^{cv,\xi,s}$ is the transition dipole moment (in low-wavenumber approximation) \cite{mkrtchian2019theory,aversa1995nonlinear}:
\begin{align}
    \mathbf d_{}^{\lambda,\bar \lambda,\xi,\xi^{\prime},s} = \left.-\mathrm ie\frac{1}{\mathcal A_{\text{UC}}}\int_{\mathcal A_{\text{UC}}}\mathrm d^2r\,u^*_{}\vphantom{u}_{\lambda,\mathbf k}^{\xi,s}(\mathbf r)\nabla_{\mathbf k}u_{\bar \lambda,\mathbf k}^{\xi^{\prime},s}(\mathbf r)\right|_{\mathbf k=\mathbf 0},
\end{align}
with unit cell area $\mathcal A_{\text{UC}}$ and bands $\lambda=c,v$, where if $\lambda = v$ then $\bar \lambda = c$ and vice versa, and the symmetry: $\left(\mathbf d_{}^{\lambda,\bar \lambda,\xi,\xi^{\prime},s}\right)^* = \mathbf d_{}^{\bar \lambda,\lambda,\xi^{\prime},\xi,s}$. This expression can be evaluated using the Bloch factors $u_{\bar \lambda,\mathbf k}^{\xi^{\prime},s}(\mathbf r)$ of the corresponding $\mathbf k\cdot\mathbf p$ Hamiltonian \cite{kormanyos2015k} yielding for $\xi=\xi^{\prime}$:
\begin{align}
    \mathbf d_{}^{c v,\xi,s}\approx\begin{cases}\dfrac{\mathrm ie\gamma\sqrt{2}}{E_{\text{gap}}^{K,K,s,s}}\mathbf e_{-},\quad\xi=K,\\
    -\dfrac{\mathrm ie\gamma\sqrt{2}}{E_{\text{gap}}^{K^{\prime},K^{\prime},s,s}}\mathbf e_{+},\quad\xi=K^{\prime},
    \end{cases}
    \label{eq:DipoleMomentExplicit}
\end{align}
with Jones vectors $\mathbf e_{\pm}$ and momentum matrix element $\gamma$ taken from Ref.~\cite{kormanyos2015k}.
The equations of motion for the pump-induced transitions in the rotating frame with pump frequency $\omega_{\text{p}}$ for normal incidence, i.e., $\mathbf E_{\mathbf Q} = \delta_{\mathbf Q,\mathbf 0}\mathbf E$, read:
\begin{align}
    \left.\mathrm i\hbar\partial_t \Poltwo{\mu}{\xi,s}\right|_{0+\text{opt}} = \left(E^{\xi,\xi,s,s}_{\mu} - \hbar\omega_{\text{p}} - \mathrm i\hbar\gamma_{\text{rad},\mu}^{\xi,s}\right)\Poltwo{\mu}{\xi,s} -\sum_{\mathbf q}\ExWFstartwo{\mathbf q}{\xi,\xi,s,s}\mathbf E^{\text{p}}\cdot\mathbf d^{cv,\xi,s}.
\end{align}
Here, $\gamma_{\text{rad},\mu}^{\xi,s} = \dfrac{\omega_{\text{p}}\left|\sum_{\mathbf q}\ExWFtwo{\mu,\mathbf q}{\xi,\xi,s,s}d^{cv,\xi,s}\right|^2}{\mathcal A2\hbar\epsilon_0c_0n_{\text{ref}}}$ is the radiative dephasing obtained by self-consistently solving Maxwell's equations \cite{knorr1996theory,jahnke1997linear}, $\mathbf E^{\text{p}}$ is the incident Gaussian pump pulse with 50~fs intensity FWHM.
The equations of motion for the
the correlations of two intravalley excitons read:
\begin{align}
    \left.\Cextwo{\mu,\nu,\mathbf Q}{\xi,\bar\xi,s,s^{\prime}}\right|_0 = \left(E_{\nu,\mathbf Q}^{\bar \xi,\bar \xi,s^{\prime},s^{\prime}} - E_{\mu,\mathbf Q}^{\xi, \xi,s,s}\right)\Cextwo{\mu,\nu,\mathbf Q}{\xi,\bar\xi,s,s^{\prime}},
\end{align}
and the equations of motion for the correlations of two intervalley excitons read:
\begin{align}
    \left.\Kextwo{\mu,\nu,\mathbf Q}{\xi,\bar \xi,\Gamma,\xi,s,s^{\prime}}\right|_0 = \left(E_{\nu,\mathbf Q}^{\Gamma,\xi,s^{\prime},s^{\prime}} - E_{\mu,\mathbf Q}^{\xi, \bar \xi,s,s}\right)\Kextwo{\mu,\nu,\mathbf Q}{\xi,\bar \xi,\Gamma,\xi,s,s^{\prime}}.
\end{align}
Regarding the excitonic occupations $\Nextwo{\mu,\mathbf Q}{\xi,\xi^{\prime},s}$, no free excitonic contributions exist.


\subsubsection{Phonon-assisted scattering}

We apply the exciton-phonon Hamiltonian \cite{selig2018dark,katzer2023exciton,brem2018exciton}, which takes into account intravalley as well as intervalley scattering of between all $K$, $K^{\prime}$, $\Lambda$, $\Lambda^{\prime}$, and $\Gamma$ valleys:
\begin{align}
\begin{split}
    \hat H_{\text{X-phon}} = &\, \sum_{\substack{\mu,\nu,\mathbf Q,\mathbf K,\alpha,\\\xi,\xi^{\prime},\xi^{\prime\prime},s,s^{\prime}}}\left(b^{\dagger}\vphantom{b}_{-\mathbf K,\alpha}^{\xi^{\prime}-\xi}+b\vphantom{b}_{\mathbf K,\alpha}^{\xi-\xi^{\prime}}\right)\left(G_{\mu,\nu,\mathbf Q+\mathbf K,\mathbf Q,\mathbf K,\alpha}^{e,s,s^{\prime},\xi^{\prime\prime},\xi,\xi^{\prime}}\Poldagtwo{\mu,\mathbf Q+\mathbf K}{\xi^{\prime\prime},\xi,s,s^{\prime}}\Poloptwo{\nu,\mathbf Q}{\xi^{\prime\prime},\xi^{\prime},s,s^{\prime}}
    -
    G_{\mu,\nu,\mathbf Q,\mathbf Q-\mathbf K,\mathbf K,\alpha}^{h,s,s^{\prime},\xi,\xi^{\prime},\xi^{\prime\prime}}\Poldagtwo{\mu,\mathbf Q}{\xi,\xi^{\prime\prime},s,s^{\prime}}\Poloptwo{\nu,\mathbf Q-\mathbf K}{\xi^{\prime}\xi^{\prime\prime},s,s^{\prime}}\right).
    \end{split}
    \label{eq:HamiltonianExcPhonInt}
\end{align}
Here, $b^{(\dagger)}\vphantom{b}_{\mathbf K,\alpha}^{\xi^{\prime}-\xi}$ is the phonon annihilation (creation) operator at phonon momentum $\mathbf K$, phonon mode $\alpha$ and phonon valley momentum $\xi^{\prime}-\xi$ and $G^{e/h}$ are the exciton-phonon matrix elements for electron ($e$) and hole ($h$) scattering:
\begin{align}
\begin{split}
    G_{\mu,\nu,\mathbf Q+\mathbf K,\mathbf Q,\mathbf K,\alpha}^{e,s,s^{\prime},\xi^{\prime\prime},\xi,\xi^{\prime\prime},\xi^{\prime}} = &\, \sum_{\mathbf q}\ExWFstartwo{\mu,\mathbf q+\beta_{\xi^{\prime\prime},\xi}^{s,s^{\prime}}(\mathbf Q+\mathbf K)}{\xi^{\prime\prime},\xi,s,s^{\prime}}\ExWFtwo{\nu,\mathbf q+\beta_{\xi^{\prime\prime},\xi^{\prime}}^{s,s^{\prime}}\mathbf Q}{\xi^{\prime\prime},\xi^{\prime},s,s^{\prime}}g_{\mathbf K,\alpha}^{c,\xi-\xi^{\prime},s^{\prime}},
    \label{eq:GOverlaps_e}
    \end{split}\\
    \begin{split}
    G_{\mu,\nu,\mathbf Q,\mathbf Q-\mathbf K,\mathbf K,\alpha}^{h,s,s^{\prime},\xi,\xi^{\prime\prime},\xi^{\prime},\xi^{\prime\prime}} = &\, \sum_{\mathbf q}\ExWFstartwo{\mu,\mathbf q-\alpha_{\xi,\xi^{\prime\prime}}^{s,s^{\prime}}\mathbf Q}{\xi,\xi^{\prime\prime},s,s^{\prime}}\ExWFtwo{\nu,\mathbf q-\alpha_{\xi^{\prime},\xi^{\prime\prime}}^{s,s^{\prime}}(\mathbf Q-\mathbf K)}{\xi^{\prime},\xi^{\prime\prime},s,s^{\prime}}g_{\mathbf K,\alpha}^{v,\ell,\xi^{\prime}-\xi,s},
    \label{eq:GOverlaps_h}
    \end{split}
\end{align}
with symmetry relations:
\begin{align}
\begin{split}
    G_{\mu,\nu,\mathbf Q+\mathbf K,\mathbf Q,\mathbf K,\alpha}^{e,s,s^{\prime},\xi,\xi^{\prime},\xi,\xi^{\prime\prime}} = &\, \left(G_{\nu,\mu,\mathbf Q,\mathbf Q+\mathbf K,-\mathbf K,\alpha}^{e,s,s^{\prime},\xi,\xi^{\prime\prime},\xi,\xi^{\prime}}\right)^*,\\
    G_{\mu,\nu,\mathbf Q,\mathbf Q-\mathbf K,\mathbf K,\alpha}^{h,s,s^{\prime},\xi,\xi^{\prime},\xi^{\prime\prime},\xi^{\prime}} = &\, \left(G_{\nu,\mu,\mathbf Q-\mathbf K,\mathbf Q,-\mathbf K,\alpha}^{h,s,s^{\prime},\xi^{\prime\prime},\xi^{\prime},\xi,\xi^{\prime}}\right)^*,\\
    g_{\mathbf K,\alpha}^{c/v,\xi-\xi^{\prime},s} = &\, \left(g_{-\mathbf K,\alpha}^{c/v,\xi^{\prime}-\xi,s}\right)^*,
    \end{split}
\end{align}
where $g_{\mathbf K,\alpha}^{c/v,\xi-\xi^{\prime},s}$ are the electron-phonon matrix elements, taken from Ref.~\cite{jin2014intrinsic} in deformation-potential approximation.
Regarding intravalley scattering, we take into account A$^{\prime}$, TO, LA and TA phonon modes.
The electron-phonon matrix elements from Ref.~\cite{jin2014intrinsic} regarding intravalley acoustic phonon scattering are rescaled by a factor of $\frac{1}{\sqrt{2}}$, which takes a vanishing piezoelectric coupling of excitons into account \cite{lengers2020theory}. Moreover, since the LO mode couples mainly via Fröhlich interaction \cite{kaasbjerg2012phonon}, its coupling to excitons for intravalley scattering vanishes due equal signs of the electron-phonon and hole-phonon Fröhlich coupling elements.
For intervalley scattering, we take A$^{\prime}$, LO, TO, LA and TA modes into account.

By applying a Markov approximation \cite{malic2013graphene} for the emerging phonon-assisted transitions and occupations and treating the phonons as a bath, we obtain closed equations on a second order Born approximation.
Excitonic transitions:
\begin{align}
    \left.\mathrm i\hbar\partial_t \Poltwo{\mu}{\xi,s}\right|_{\text{X-phon}} = - \mathrm i\hbar\gamma_{\text{phon},\mu}^{\xi,s}(\hbar\omega_{\text{p}})\Poltwo{\mu}{\xi,s}.
    \label{eq:P_Phon}
\end{align}
The phonon-assisted dephasing $\gamma_{\text{phon},\mu}^{\xi,s}$ is calculated self-consistently \cite{lengers2020theory} in a rotating frame, cf.\ Eq.~\eqref{eq:DephasingPhon}.
Excitonic occupations:
\begin{align}
\begin{split}
    \left.\partial_t\Nextwo{\mu,\mathbf Q}{\xi,\xi^{\prime},s}\right|_{\text{X-phon}} = &\,
    \sum_{\substack{\nu,\xi^{\prime\prime},\xi^{\prime\prime\prime}}}\Gamma^{\text{form},s,s,\xi^{\prime\prime},\xi^{\prime\prime},\xi,\xi^{\prime}}_{\nu,\mu,\mathbf Q}
    \left|\Poltwo{\nu}{\xi^{\prime\prime},s}\right|^2\\
    &\,+\sum_{\substack{\nu,\mathbf K,\xi^{\prime\prime},\xi^{\prime\prime\prime}}}\left(\Gamma^{\text{in},s,s,\xi^{\prime\prime},\xi^{\prime\prime\prime},\xi,\xi^{\prime}}_{\nu,\mu,\mathbf K,\mathbf Q,-\mathbf Q+\mathbf K}
    \Nextwo{\nu,\mathbf K}{\xi^{\prime\prime},\xi^{\prime\prime\prime},s} - \Gamma^{\text{out},s,s,\xi^{\prime\prime},\xi^{\prime\prime\prime},\xi,\xi^{\prime}}_{\nu,\mu,\mathbf K,\mathbf Q,-\mathbf Q+\mathbf K}
    \Nextwo{\mu,\mathbf Q}{\xi,\xi^{\prime},s}\right).
    \end{split}
    \label{eq:Boltzmann_Equation}
\end{align}
Correlations of two intravalley excitons:
\begin{align}
    \left.\partial_t\Cextwo{\mu,\nu,\mathbf Q}{\xi,\bar\xi,s,s^{\prime}}\right|_{\text{X-phon}} = - \frac{1}{2}\sum_{\substack{\nu,\mathbf K,\xi^{\prime\prime},\xi^{\prime\prime\prime}}}\left(
    \Gamma^{\text{out},s,s,\xi^{\prime\prime},\xi^{\prime\prime\prime},\xi,\xi}_{\nu,\mu,\mathbf K,\mathbf Q,-\mathbf Q+\mathbf K} + \Gamma^{\text{out},s^{\prime},s^{\prime},\xi^{\prime\prime},\xi^{\prime\prime\prime},\bar \xi,\bar \xi}_{\nu,\mu,\mathbf K,\mathbf Q,-\mathbf Q+\mathbf K}
    \right)\Cextwo{\mu,\nu,\mathbf Q}{\xi,\bar\xi,s,s^{\prime}}.
    \label{eq:C_PhononEquations}
\end{align}
Correlations of two intervalley excitons:
\begin{align}
    \left.\partial_t\Kextwo{\mu,\nu,\mathbf Q}{\xi,\bar \xi,\Gamma,\xi,s,s^{\prime}}\right|_{\text{X-phon}} = 
    - \frac{1}{2}\sum_{\substack{\nu,\mathbf K,\xi^{\prime\prime},\xi^{\prime\prime\prime}}}\left(
    \Gamma^{\text{out},s,s,\xi^{\prime\prime},\xi^{\prime\prime\prime},\xi,\bar \xi}_{\nu,\mu,\mathbf K,\mathbf Q,-\mathbf Q+\mathbf K} + \Gamma^{\text{out},s^{\prime},s^{\prime},\xi^{\prime\prime},\xi^{\prime\prime\prime},\Gamma,\xi}_{\nu,\mu,\mathbf K,\mathbf Q,-\mathbf Q+\mathbf K}
    \right)
    \Kextwo{\mu,\nu,\mathbf Q}{\xi,\bar \xi,\Gamma,\xi,s,s^{\prime}}.
    \label{eq:K_PhononEquations}
\end{align}
Here, exciton-phonon interaction causes the formation of incoherent excitonic occupations $\Nextwo{\mu,\mathbf Q}{\xi,\xi^{\prime},s}$ by coherent transitions $\Poltwo{\mu,\mathbf Q}{\xi,s}$ (first line in Eq.~\eqref{eq:Boltzmann_Equation}) and a scattering between incoherent excitonic occupations (second line in Eq.~\eqref{eq:Boltzmann_Equation}) as well as a decay of the two-exciton correlations $\Cextwo{\mu,\nu,\mathbf Q}{\xi,\bar\xi,s,s^{\prime}}$ and $\Kextwo{\mu,\nu,\mathbf Q}{\xi,\bar \xi,s,s^{\prime}}$ in Eq.~\eqref{eq:C_PhononEquations} and Eq.~\eqref{eq:K_PhononEquations}. Phonon-assisted scattering between the respective two-exciton correlations $\Cextwo{\mu,\nu,\mathbf Q}{\xi,\bar\xi,s,s^{\prime}}$ or $\Kextwo{\mu,\nu,\mathbf Q}{\xi,\bar \xi,s,s^{\prime}}$ are not possible, since we allow only one-phonon processes.
The outscattering rates appearing in Eq.~\eqref{eq:Boltzmann_Equation}, Eq.~\eqref{eq:C_PhononEquations} and Eq.~\eqref{eq:K_PhononEquations} read:
\begin{align}
\begin{split}
    &\Gamma^{\text{out},s,s,\xi^{\prime\prime},\xi^{\prime\prime\prime},\xi,\xi^{\prime}}_{\nu,\mu,\mathbf K,\mathbf Q,-\mathbf Q+\mathbf K} = \\
    & \frac{2\pi}{\hbar}\sum_{\pm,\alpha}\left|G^{e,s,s,\xi,\xi^{\prime},\xi,\xi^{\prime}}_{\nu,\mu,\mathbf K,\mathbf Q,\mathbf Q-\mathbf K,\alpha}
    + G^{h,s,s,\xi,\xi^{\prime},\xi,\xi^{\prime}}_{\nu,\mu,\mathbf K,\mathbf Q,-\mathbf Q+\mathbf K,\alpha}
    \right|^2\left(\frac{1}{2}\pm\frac{1}{2}+ n_{\pm\mathbf Q\mp\mathbf K,\alpha}^{\Gamma}\right)
    \delta \left(E_{\nu,\mathbf K}^{\xi,\xi^{\prime},s,s}-E_{\mu,\mathbf Q}^{\xi,\xi^{\prime},s,s}\pm\hbar\Omega_{\pm\mathbf Q\mp\mathbf K,\alpha}^{\Gamma}\right)\delta_{\xi^{\prime\prime},\xi}^{\xi^{\prime\prime\prime},\xi^{\prime}}\\
    & + \frac{2\pi}{\hbar}\sum_{\pm,\alpha}\left|G^{e,s,s,\xi,\xi^{\prime\prime\prime},\xi^{\prime}}_{\nu,\mu,\mathbf K,\mathbf Q,\mathbf Q-\mathbf K,\alpha}\right|^2\left(\frac{1}{2}\pm\frac{1}{2}+ n_{\pm\mathbf Q\mp\mathbf K,\alpha}^{\mp\xi^{\prime\prime\prime}\pm\xi^{\prime}}\right)\delta \left(E_{\nu,\mathbf K}^{\xi^{\prime\prime},\xi^{\prime\prime\prime},s,s}-E_{\mu,\mathbf Q}^{\xi,\xi^{\prime},s,s}\pm\hbar\Omega_{\pm\mathbf Q\mp\mathbf K,\alpha}^{\mp\xi^{\prime\prime\prime}\pm\xi^{\prime}}\right)\delta_{\xi^{\prime\prime},\xi}\\
    & +\frac{2\pi}{\hbar}\sum_{\pm,\alpha}\left|G^{h,s,s,\xi^{\prime\prime},\xi,\xi^{\prime}}_{\nu,\mu,\mathbf K,\mathbf Q,-\mathbf Q+\mathbf K,\alpha}\right|^2\left(\frac{1}{2}\pm\frac{1}{2}+ n_{\pm\mathbf Q\mp\mathbf K,\alpha}^{\pm\xi^{\prime\prime}\mp\xi}\right)\delta \left(E_{\nu,\mathbf K}^{\xi^{\prime\prime},\xi^{\prime},s,s}-E_{\mu,\mathbf Q}^{\xi,\xi^{\prime},s,s}\pm\hbar\Omega_{\pm\mathbf Q\mp\mathbf K,\alpha}^{\pm\xi^{\prime\prime}\mp\xi}\right)\delta_{\xi^{\prime\prime\prime},\xi^{\prime}}.
    \end{split}
    \label{eq:ScatteringRates}
\end{align}
Here, the first term denotes intravalley electron and hole scattering, the second denotes intervalley electron scattering and the third term denotes intervalley hole scattering. $\hbar\Omega_{\pm\mathbf Q\mp\mathbf K,\alpha}^{\pm\xi^{\prime}\mp\xi}$ is the phonon dispersion with regard to the momentum $\pm\mathbf Q\mp\mathbf K$ around the valley momentum $\pm\xi^{\prime}\mp\xi$ and mode $\alpha$. $n_{\pm\mathbf Q\mp\mathbf K,\alpha}^{\mp\xi^{\prime}\pm\xi}$ is the phonon occupation, which is assumed as a static Bose-Einstein distribution in the bath approximation.

The scattering rates are obtained via:
\begin{align}
    \Gamma^{\text{in},s,s,\xi,\xi^{\prime\prime},\xi^{\prime}}_{\nu,\mu,\mathbf K,\mathbf Q,-\mathbf Q+\mathbf K} = \Gamma^{\text{out},s,s,\xi,\xi^{\prime},\xi^{\prime\prime}}_{\mu,\nu,\mathbf Q,\mathbf K,\mathbf Q-\mathbf K}.
\end{align}
The phonon-assisted dephasing $\gamma_{\text{phon},\mu}^{\xi,s}$ in Eq.~\eqref{eq:P_Phon} taking into account a finite coherence decay of higher correlations \cite{lengers2020theory} is obtained by self-consistently solving the following equation:
\begin{align}
    \gamma_{\text{phon},\mu}^{\xi,s} = \left.\frac{1}{2}\sum_{\nu,\mathbf K,\xi^{\prime\prime},\xi^{\prime\prime\prime}}\Gamma^{\text{out},s,s,\xi^{\prime\prime},\xi^{\prime\prime\prime},\xi,\xi}_{\nu,\mu,\mathbf K,\mathbf 0,\mathbf K}\right|_{\substack{E_{\nu}^{\xi,\xi,s,s} = \hbar\omega_{\text{p}},~\delta\rightarrow \mathcal L_{\gamma}}},
    \label{eq:DephasingPhon}
\end{align}
where the Dirac delta function in $\Gamma^{\text{out}}_{}$ is replaced by a Lorentzian, which takes into account a finite lifetime of higher correlations \cite{lengers2020theory}:
\begin{align}
    \delta(E_{\mathbf Q}-\hbar\omega_p\pm\hbar \Omega_{\mathbf Q})\rightarrow \mathcal L_{\gamma_{\text{phon},\mu}^{\xi,s}}(E_{\mathbf Q}-\hbar\omega_p\pm\hbar\Omega_{\mathbf Q}) = \frac{1}{\pi}\frac{\hbar\gamma_{\text{phon},\mu}^{\xi,s}}{(E_{\mathbf Q}-\hbar\omega_{\text{p}}\pm\hbar \Omega_{\mathbf Q})^2 + (\hbar\gamma_{\text{phon},\mu}^{\xi,s})^2}.
\end{align}
Similarly, the formation rate $\Gamma^{\text{form}}_{}$ appearing in Eq.~\eqref{eq:Boltzmann_Equation} is evaluated via:
\begin{align}
    \Gamma^{\text{form},s,s,\xi^{\prime\prime},\xi^{\prime\prime},\xi,\xi^{\prime}}_{\nu,\mu,\mathbf Q} = \left.
    \Gamma^{\text{in},s,s,\xi^{\prime\prime},\xi^{\prime\prime},\xi,\xi^{\prime}}_{\nu,\mu,\mathbf 0,\mathbf Q,-\mathbf Q}\right|_{\substack{E_{\nu}^{\xi^{\prime\prime},\xi^{\prime\prime},s,s} = \hbar\omega_{\text{p}},~\delta\rightarrow \mathcal L_{\gamma}}}.
\end{align}


\subsubsection{Exchange interaction}
The Coulomb exchange Hamiltonian (energy transfer) reads:
\begin{align}
    \hat H_{\text{coul,X}}= \sum_{\substack{\mathbf k,\mathbf k^{\prime},\mathbf q,\\\mathbf G,\\\xi,\xi^{\prime},\xi^{\prime\prime},s,s^{\prime}}} V_{\mathbf q+ \mathbf K^{\xi^{\prime\prime}}+\mathbf G}
    F_{\mathbf k,\mathbf q+\mathbf G}^{c,v,\xi+\xi^{\prime\prime},\xi,s}F_{\mathbf k^{\prime},-\mathbf q-\mathbf G}^{v,c,\xi^{\prime}-\xi^{\prime\prime},\xi^{\prime},s^{\prime}}
    \cdagtwo{\mathbf k+\mathbf q}{\xi+\xi^{\prime\prime},s}\vdagtwo{\mathbf k^{\prime}-\mathbf q}{\xi^{\prime}-\xi^{\prime\prime},s^{\prime}}\cndagtwo{\mathbf k^{\prime}}{\xi^{\prime},s^{\prime}}\vndagtwo{\mathbf k}{\xi,s},
    \label{eq:Coulomb_exchange_hamiltonian_general}
\end{align}
with form factors:
\begin{align}
    F_{\mathbf k,\mathbf q+\mathbf G}^{\lambda,\lambda^{\prime},\xi,\xi^{\prime},s} = \frac{1}{\mathcal A_{\text{UC}}}\int_{\mathcal A_{\text{UC}}}\mathrm d^2r\,{u^*}_{\lambda,\mathbf k+\mathbf q}^{\xi,s}(\mathbf r)\mathrm e^{\mathrm i\mathbf G\cdot\mathbf r}u_{\lambda^{\prime},\mathbf k}^{\xi^{\prime},s}(\mathbf r).
\end{align}
Here, we use a macroscopically screened Coulomb potential $V_{\mathbf q+ \mathbf K^{\xi^{\prime\prime}}+\mathbf G}$ with $\mathbf G=\mathbf G^{\prime}$. In fact, the exchange potential is unscreened \cite{rohlfing2000electron,qiu2015nonanalyticity}, but an interaction with all other bands introduces an overall attenuation of the bare exchange potential. Hence, assuming the exchange potential as macroscopically screened in our two-band effective-mass approach serves as to not overestimate the exchange interaction strength.

In Eq.~\eqref{eq:Coulomb_exchange_hamiltonian_general}, we distinguish between the scattering between intravalley excitons ($\xi^{\prime\prime}=0$) and between intervalley excitons ($\xi^{\prime\prime}\neq 0$).

(i) Intravalley excitons:
\begin{align}
    \hat H_{\text{coul,X,intra}}= \sum_{\substack{\mathbf k,\mathbf k^{\prime},\mathbf q,\mathbf G,\\\xi,\xi^{\prime},s,s^{\prime}}} V^{}_{\mathbf q+\mathbf G}
    F_{\mathbf k,\mathbf q+\mathbf G}^{c,v,\xi,\xi,s}F_{\mathbf k^{\prime},-\mathbf q-\mathbf G^{\prime}}^{v,c,\xi^{\prime},\xi^{\prime},s^{\prime}}
    \cdagtwo{\mathbf k+\mathbf q}{\xi,s}\vdagtwo{\mathbf k^{\prime}-\mathbf q}{\xi^{\prime},s^{\prime}}\cndagtwo{\mathbf k^{\prime}}{\xi^{\prime},s^{\prime}}\vndagtwo{\mathbf k}{\xi,s},
\end{align}

The coupling element can be evaluated in low-wavenumber approximation and reads:
\begin{align}
    V_{\text{X},\mathbf q}^{\xi,\xi^{\prime},s,s^{\prime}} = V_{\text{SR}}\delta_{\xi,\xi^{\prime}}\delta_{s,s^{\prime}} +  V_{\mathbf q}^{}\frac{1}{e^2}\left(\mathbf q\cdot \mathbf d_{}^{vc,\xi,s}\vphantom{d_{}^{cv,\xi^{\prime},s^{\prime}}}\right)\left(\mathbf q\cdot \mathbf d_{}^{cv,\xi^{\prime},s^{\prime}}\right)\delta_{\mathbf G,\mathbf 0}.
    \label{eq:exchange_potential}
\end{align}
The first term is the local short-range interaction with non-vanishing reciprocal lattice vector $\mathbf G\neq\mathbf 0$, which occurs only as intravalley interaction, and the second is the local- ($\xi=\xi^{\prime}$) and non-local ($\xi\neq\xi^{\prime}$) long-range interaction with vanishing reciprocal lattice vector $\mathbf G=\mathbf 0$. The long-range interaction can occur via non-resonant double spin flips within equal valleys \cite{guo2019exchange}, via spin-conserved non-resonant scattering between distinct valleys and via resonant double spin flip scattering between distinct valleys. The latter interaction is also known as Maialle-Silva-Sham mechanism \cite{combescot2023ab,selig2019ultrafast,yu2014valley,maialle1993exciton}. 

In Eq.~\ref{eq:exchange_potential}, $\mathbf d_{}^{vc,\xi,s}$ is the transition dipole moment, cf.~Eq.~\eqref{eq:DipoleMomentExplicit}, hence the long-range exchange interaction can be viewed as a Coulomb dipole-dipole interaction.

The exchange Hamiltonian in the excitonic picture reads:
\begin{align}
    \hat H_{\text{coul,X,intra}} =  \sum_{\substack{\mu,\nu,\mathbf Q,\\\xi,\xi^{\prime},s,s^{\prime}}}X_{\text{intra},\mu,\nu,\mathbf Q}^{\xi,\xi^{\prime},s,s^{\prime}}\Poldagtwo{\mu,\mathbf Q}{\xi,\xi,s,s}\Poloptwo{\nu,\mathbf Q}{\xi^{\prime},\xi^{\prime},s^{\prime},s^{\prime}},
    \label{eq:HamiltonianExchange_ExcitonPicture}
\end{align}
with matrix element:
\begin{align}
    X_{\text{intra},\mu,\nu,\mathbf Q}^{\xi,\xi^{\prime},s,s^{\prime}} = V_{\text{X},\mathbf Q}^{\xi^{\prime},\xi,s^{\prime},s} 
    \sum_{\mathbf q}
    {\varphi^*}_{\mu,\mathbf q}^{\xi,\xi,s,s}\sum_{\mathbf q^{\prime}}\varphi_{\nu,\mathbf q^{\prime}}^{\xi^{\prime},\xi^{\prime},s^{\prime}, s^{\prime}}.
    \label{eq:ExchangeMatrixElement}
\end{align}
The short-range contribution in Eq.~\eqref{eq:HamiltonianExchange_ExcitonPicture} causes a blue shift of the optical bright excitonic transition and entails a calculation via first-principle methods due to the form factors at \mbox{$\mathbf G\neq\mathbf 0$} \cite{qiu2015nonanalyticity}. However, it is possible to obtain an estimation of the magnitude of the short-range exchange potential $V_{\text{SR}}$ via the experimentally measured excitonic bright-dark splitting as follows: 
The lowest energy transition in a 1L-WSe$_2$ is optically dark, i.e.,~the spin-unlike exciton is located around $47$~meV \cite{zhang2017, molas2017} (on a SiO$_2$ substrate)
below the optically addressable spin-like exciton.
Within our effective mass approach, we obtain binding energies of the lowest spin-like and lowest spin-unlike exciton of $395.5$~meV and $427.4$~meV for a 1L-WSe$_2$ on a SiO$_2$ substrate, cf.\ Eq.~\eqref{eq:WannierEquation}. In combination with the conduction band splitting of 13~meV (mean value of Refs.~\cite{ren2023measurement,kapuscinski2021rydberg}) we estimate the short-range intravalley exchange as $\Delta_{\text{X}} = 5.2$~meV.
Note, that while the effective masses from DFT in Ref.~\cite{kormanyos2015k} are in agreement with measurements \cite{le2015spin}, the spin-orbit conduction band splitting of 37~meV obtained by DFT in Ref.~\cite{kormanyos2015k} would yield a negative short-range intravalley exchange energy shift, i.e.,\ a red shift, which would be unphysical. 
Recent DFT calculations from Ref.~\cite{boccuni2024unveiling} indicate, that standard DFT \cite{kormanyos2015k} systematically overestimates (underestimates) the conduction (valence) band spin splitting in WSe$_2$ monolayers: Their calculations show, that Breit interaction \cite{breit1929effect}, which is usually not considered, strongly reduces the discrepancy between theory and experiment. The remaining few meV difference can be explained via, e.g, strain, which is known to be always present in TMDs on substrates to some degree resulting in a reduction of the conduction band spin splitting compared to an unstrained TMD \cite{sahu2024strain,yang2023strain}.
The short-range intravalley exchange blue shift of the $1s$ A exciton reads \cite{katsch2020exciton}:
\begin{align}
    \Delta_{\text{X},1s}^{K,K,\Uparrow,\uparrow} = V_{\text{SR}}^{K,K}\left|\sum_{\mathbf q}\ExWFtwo{1s,\mathbf q}{K,K,\Uparrow,\uparrow}\right|^2 = \Delta_{\text{X},1s}^{K^{\prime},K^{\prime},\Downarrow,\downarrow}.
\end{align}
which enables the estimation of the short-range exchange potential of intravalley excitons $V_{\text{SR}}$, from which the short-range intravalley blue shift of the $1s$ B exciton can be calculated:
\begin{align}
    \Delta_{\text{X},1s}^{K,K,\Downarrow,\downarrow} = V_{\text{SR}}\left|\sum_{\mathbf q}\ExWFtwo{1s,\mathbf q}{K,K,\Downarrow,\downarrow}\right|^2 = \Delta_{\text{X},1s}^{K^{\prime},K^{\prime},\Uparrow,\uparrow}.
\end{align}

(ii) Intervalley excitons:
\begin{align}
    \hat H_{\text{coul,X,inter}}= \sum_{\substack{\mathbf k,\mathbf k^{\prime},\mathbf q,\mathbf G,\\\xi,\xi^{\prime\prime} (\xi^{\prime\prime}\neq 0),s}}
    V_{\mathbf q+ \mathbf K^{\xi^{\prime\prime}}+\mathbf G}
    F_{\mathbf k,\mathbf q+\mathbf G}^{cv,\xi+\xi^{\prime\prime},\xi,s,s} F_{\mathbf k^{\prime},-\mathbf q-\mathbf G}^{vc,\xi,\xi+\xi^{\prime\prime},s,s}
    \cdagtwo{\mathbf k+\mathbf q}{\xi+\xi^{\prime\prime},s}\vdagtwo{\mathbf k^{\prime}-\mathbf q}{\xi,s}\cndagtwo{\mathbf k^{\prime}}{\xi+\xi^{\prime\prime},s}\vndagtwo{\mathbf k}{\xi,s},.
\end{align}
Here, we already restricted the interaction processes to equal valleys and spins.
In the excitonic picture, we obtain:

\begin{align}
    \hat H_{\text{coul,X,inter}} =  \sum_{\substack{\mu,\nu,\mathbf Q,\xi,s}}X_{\text{inter},\mu,\nu}^{\xi,\bar \xi,s,s}\Poldagtwo{\mu,\mathbf Q}{\xi,\bar \xi,s,s}\Poloptwo{\nu,\mathbf Q}{\xi,\bar \xi,s,s},
    \label{eq:HamiltonianExchange_ExcitonPicture_Intervalley}
\end{align}
with matrix element:
\begin{align}
    X_{\text{inter},\mu,\nu}^{\xi,\bar \xi,s,s} = V_{\text{X},\mathbf K^{\xi}-\mathbf K^{\bar \xi}}^{\xi^{\prime},\xi,s^{\prime},s} 
    \sum_{\mathbf q}
    {\varphi^*}_{\mu,\mathbf q}^{\xi,\bar \xi,s,s}\sum_{\mathbf q^{\prime}}\varphi_{\nu,\mathbf q^{\prime}}^{\xi,\bar \xi,s, s}.
    \label{eq:ExchangeMatrixElementIntervalley}
\end{align}

It has been experimentally and theoretically shown, that short-range intravalley exchange also affects intervalley excitons \cite{li2022intervalley,yang2022relaxation,he2020valley}. Since the short-range exchange potential of intravalley excitons $V_{\text{SR}}$ involves evaluating the Coulomb potential at a reciprocal lattice vector $\mathbf G$ with $V_{\mathbf G}$, we can approximate the short-range exchange potential of intervalley excitons via: $V_{\text{SR}}^{K,K^{\prime}}\approx V_{\text{SR}}^{KK}\frac{V_{\mathbf K}}{V_{\mathbf G}}$, as the short-range exchange interaction of intervalley $K$-$K^{\prime}$ ($K^{\prime}$-$K$) excitons involves evaluating the Coulomb potential at the valley momentum $\mathbf K$. An analog procedure is done for the intervalley excitons involving the $\Lambda$/$\Lambda^{\prime}$ and $\Gamma$ valleys, which, in particular, removes the energy-degeneracy of, e.g., spin-bright \mbox{$\Gamma_{\Uparrow}$-$K_{\uparrow}$} and spin-dark \mbox{$\Gamma_{\Downarrow}$-$K_{\uparrow}$} excitons, cf.\ Fig.~\ref{fig:ExcitonicEnergies}.

The long-range contribution in Eq.~\eqref{eq:HamiltonianExchange_ExcitonPicture} leads to intervalley coupling of intravalley excitonic occupations via:
\begin{align}
\label{eq:N_Equations_Exchange}
    \left.\mathrm i\hbar\partial_t {N_{}^{}}_{\mu,\mathbf Q}^{\xi,\xi,s}\right|_{\text{coul,X}} = 2\mathrm i\sum_{\nu}\text{Im}\left(X_{\mu,\nu,\mathbf Q}^{\xi,\bar \xi,s,\bar s}{C_{}^{}}_{\mu,\nu,\mathbf Q}^{\xi,\bar \xi,s,\bar s}\right).
\end{align}
Here, the intravalley occupations $\Nextwo{\mu,\mathbf Q}{\xi,\xi,s}$ couple to intervalley correlations $\Cextwo{\mu,\mathbf Q}{\xi,\bar \xi,s,\bar s}$ composed of two intravalley excitons at opposite valleys and spin, whose equations of motion read:
\begin{align}
\label{eq:C_Equations_Exchange}
    \left.\mathrm i\hbar\partial_t {C_{}^{}}_{\mu,\nu,\mathbf Q}^{\xi,\bar \xi, s,\bar s}\right|_{\text{coul,X}} = 
    X_{\nu,\mu,\mathbf Q}^{\bar \xi,\xi,\bar s, s}\left({N_{}^{}}_{\mu,\mathbf Q}^{\xi,\xi,s,s}-{N_{}^{}}_{\nu,\mathbf Q}^{\bar \xi,\bar \xi,\bar s,\bar s}\right),
\end{align}
which then couple back to intravalley occupations at opposite valleys leading to an occupation transfer of intravalley occupations between the $K$ and $K^{\prime}$ valleys.


\subsubsection{Dexter interaction}

We denote the intraband Coulomb interaction with large momentum transfer as Dexter interaction (charge transfer) \cite{berghauser2018, bernal2018exciton}. The corresponding Hamiltonian reads:
\begin{align}
    \hat H_{\text{coul,Dex}}= \sum_{\substack{\mathbf k,\mathbf k^{\prime},\mathbf q,\mathbf G,\mathbf G^{\prime},\\\xi,\xi^{\prime},\xi^{\prime\prime} (\xi^{\prime\prime}\neq 0),\\s,s^{\prime}}}
    V_{\mathbf q+ \mathbf K^{\xi^{\prime\prime}},\mathbf G,\mathbf G^{\prime}}
    F_{\mathbf k,\mathbf q+\mathbf G}^{c,c,\xi+\xi^{\prime\prime},\xi}
    F_{\mathbf k^{\prime},-\mathbf q-\mathbf G^{\prime}}^{v,v,\xi^{\prime}-\xi^{\prime\prime},\xi^{\prime}}
    \cdagtwo{\mathbf k+\mathbf q+\mathbf G}{\xi+\xi^{\prime\prime},s}\vdagtwo{\mathbf k^{\prime}-\mathbf q-\mathbf G^{\prime}}{\xi^{\prime}-\xi^{\prime\prime},s^{\prime}}\vndagtwo{\mathbf k^{\prime}}{\xi^{\prime},s^{\prime}}\cndagtwo{\mathbf k}{\xi,s},
    \label{eq:CoulombHamiltonianDexterElectronHole}
\end{align}
where $V_{\mathbf q,\mathbf G,\mathbf G^{\prime}}$ is the screened Coulomb potential containing non-local contributions with \mbox{$\mathbf G\neq\mathbf G^{\prime}$} \cite{rohlfing2000electron}. For \mbox{$\mathbf G=\mathbf G^{\prime}$}, the Coulomb potential can be directly linked to the macroscopically screened Coulomb potential $V_{\mathbf q}$ via \mbox{$V_{\mathbf q,\mathbf G,\mathbf G} = V_{\mathbf q+\mathbf G}$}.
In contrast to the double-spin flip exchange interactions as described before, Dexter interaction leads to a spin-conserved intervalley interaction.

\begin{figure}[ht]
    \centering
    \includegraphics[width = 1.0\linewidth]{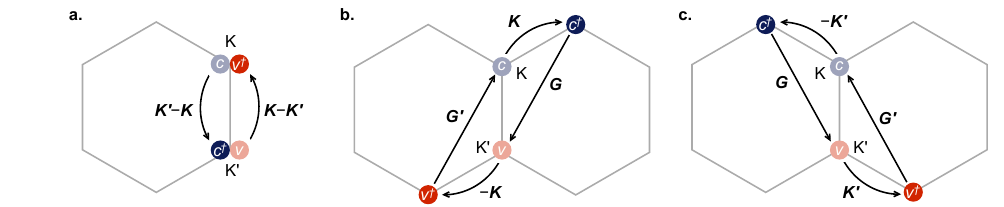}
    \caption{
    Dexter scattering in reciprocal space between intravalley excitons visualized in conduction (blue) and valence (red) electrons. Here, $\hat v/\hat c^{(\dagger)}$ annihilates (creates) an electron in the valence/conduction band, cf.~Eq.~\eqref{eq:CoulombHamiltonianDexterElectronHole}. 
    Panels \textbf{a}--\textbf{c} show possible scattering geometries involving the $K$ and $K^{\prime}$ valleys. In Fig.~\ref{fig:Dexter_intravalley_intravalley_exciton}(\textbf{a}), the corresponding process is visualized in electron-hole pairs.
    }
    \label{fig:Dexter_intravalley_intravalley_el}
\end{figure}

In our case, we identify two relevant scattering channels:

(i) Dexter interaction between intravalley excitons, where the individual holes and electrons scatter between distinct $K$/$K^{\prime}$ valleys, which involves one normal (\mbox{$\mathbf G=\mathbf 0$} and \mbox{$\mathbf G^{\prime}=\mathbf 0$}) and two Umklapp processes (\mbox{$\mathbf G=\mathbf G^{\prime} \neq \mathbf 0$}) with Bloch form factors 
$F_{\mathbf k,\mathbf q}^{c,c,K^{\prime},K,s}$, $F_{\mathbf k^{\prime},-\mathbf q}^{v,v,K,K^{\prime},s}$
and $F_{\mathbf k,\mathbf q+\mathbf G}^{c,c,K^{\prime},K,s}$, $F_{\mathbf k^{\prime},-\mathbf q-\mathbf G}^{v,v,K,K^{\prime},s}$. 
Here, the Coulomb potential can be described by a macroscopically screened Coulomb potential \mbox{$V_{\mathbf q+ \mathbf K^{\xi^{\prime\prime}},\mathbf G,\mathbf G} = V_{\mathbf q+ \mathbf K^{\xi^{\prime\prime}}+\mathbf G}$}. 
In Fig.~\ref{fig:Dexter_intravalley_intravalley_el}, we depict the one normal and two Umklapp processes visualized in conduction/valence band electrons, cf.~Eq.~\eqref{eq:CoulombHamiltonianDexterElectronHole}, in reciprocal space. In Fig.~\ref{fig:Dexter_intravalley_intravalley_exciton}(\textbf{a}), we depict the combined process visualized in electron-hole pairs in a simplified band scheme.

\begin{figure}[ht]
    \centering
    \includegraphics[width = 1.0\linewidth]{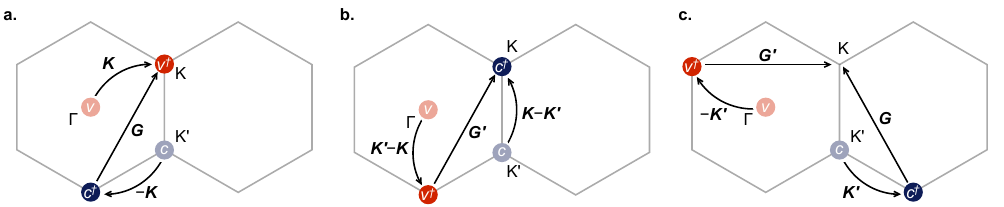}
    \caption{
    Dexter scattering in reciprocal space between intervalley excitons visualized in conduction (blue) and valence (red) electrons. Here, $\hat v/\hat c^{(\dagger)}$ annihilates (creates) an electron in the valence/conduction band, cf.~Eq.~\eqref{eq:CoulombHamiltonianDexterElectronHole}. Panels \textbf{a}--\textbf{c} show possible scattering geometries involving the $\Gamma$, $K$ and $K^{\prime}$ valleys. In Fig.~\ref{fig:Dexter_intravalley_intravalley_exciton}(\textbf{b}), the corresponding process is visualized in electron-hole pairs.
    }
    \label{fig:Dexter_intervalley_intervalley_el}
\end{figure}

\begin{figure}[ht]
    \centering
    \includegraphics[width = 1.0\linewidth]{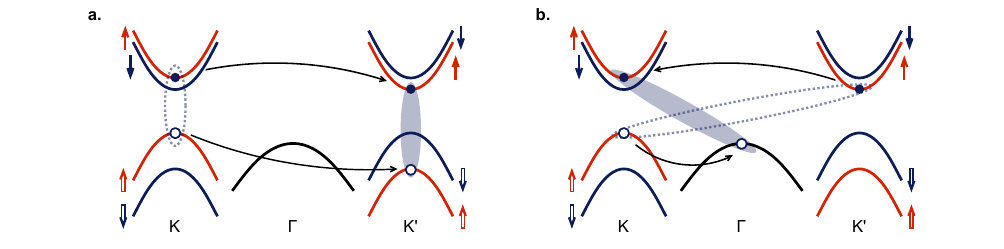}
    \caption{
    The combined process from Fig.~\ref{fig:Dexter_intravalley_intravalley_el} and from Fig.~\ref{fig:Dexter_intervalley_intervalley_el} visualized in terms of electron-hole pairs.
    }
    \label{fig:Dexter_intravalley_intravalley_exciton}
\end{figure}

(ii) Dexter interaction involving intervalley excitons, cf.\ Fig.~\ref{fig:Dexter_intervalley_intervalley_el} and Fig.~\ref{fig:Dexter_intravalley_intravalley_exciton}b, where the holes scatter between $K$ and $\Gamma$ valleys and the electrons between $K$ and $K^{\prime}$ valleys, which involves three Umklapp processes with \mbox{$\mathbf G\neq\mathbf G^{\prime}$} and Bloch form factors $F_{\mathbf k,\mathbf q+\mathbf G}^{c,c,K,K^{\prime},s}$ 
$F_{\mathbf k^{\prime},-\mathbf q-\mathbf G^{\prime}}^{v,v,K,\Gamma,s}$. 
Since \mbox{$\mathbf G\neq\mathbf G^{\prime}$}, this scattering process cannot be related to a macroscopically screened Coulomb potential as in (i). 
The evaluation of the Bloch form factors between distinct valleys and the non-local screened Coulomb potential $V_{\mathbf q,\mathbf G,\mathbf G^{\prime}}$ needs \textit{ab-initio} methods \cite{rohlfing2000electron}, which are beyond the scope of this work. Thus, to proceed, we combine all three processes (normal and Umklapp) into a single process with Bloch form factors of unity and a macroscopically screened Coulomb potential at a large valley momentum \mbox{$V_{\mathbf q+\mathbf K^{\xi} + \mathbf G} = V_{\mathbf q+\mathbf K^{\xi},\mathbf G,\mathbf G^{\prime}} \equiv V_{\mathbf K}$}, as \mbox{$\mathbf q\ll\mathbf K^{\xi}$}. This rough approximation already yields good agreement with experiments in the case of Dexter interaction of intravalley excitons in process (i) \cite{timmer2024ultrafast,bernal2018exciton} and, hence, we assume a similar behavior for the Dexter interaction between intervalley excitons in (ii). 
In Fig.~\ref{fig:Dexter_intervalley_intervalley_el}, we depict the three Umklapp processes visualized in conduction/valence band electrons, cf.~Eq.~\eqref{eq:CoulombHamiltonianDexterElectronHole}, in reciprocal space. In Fig.~\ref{fig:Dexter_intravalley_intravalley_exciton}(\textbf{b}), we depict the combined process visualized in electron-hole pairs in a simplified band scheme.



In the excitonic picture, the Dexter Hamiltonian reads:
\begin{align}
\begin{split}
    \hat H_{\text{coul,Dex}}= &\, \sum_{\substack{\mu,\nu,\mathbf Q,\\\xi,\xi^{\prime},\xi^{\prime\prime},(\xi^{\prime\prime}\neq 0),\\s,s^{\prime}}}
    D_{\mu,\nu,\mathbf K^{\xi^{\prime\prime}}}^{
    \xi^{\prime},\xi+\xi^{\prime\prime},\xi^{\prime}-\xi^{\prime\prime},\xi,s,s^{\prime}
    ,s,s^{\prime}}
    \Poldagtwo{\mu,\mathbf Q}{\xi^{\prime},\xi+\xi^{\prime\prime},s,s^{\prime}}
    \Poloptwo{\nu,\mathbf Q}{\xi^{\prime}-\xi^{\prime\prime},\xi,s,s^{\prime}},
    \end{split}
\end{align}
with matrix element:
\begin{align}
    D_{\mu,\nu,\mathbf K^{\xi^{\prime\prime}}}^{
    \xi^{\prime},\xi+\xi^{\prime\prime},\xi^{\prime}-\xi^{\prime\prime},\xi,s,s^{\prime}
    ,s,s^{\prime}} = - V^{}_{\mathbf K^{\xi^{\prime\prime}}}\sum_{\mathbf q}\ExWFstartwo{\mu,\mathbf q}{\xi^{\prime},\xi+\xi^{\prime\prime},s,s^{\prime}}\sum_{\mathbf q^{\prime}}\ExWFtwo{\nu,\mathbf q^{\prime}}{\xi^{\prime}-\xi^{\prime\prime},\xi,s,s^{\prime}}.
    \label{eq:DexterMatrixElement}
\end{align}

The equations of motion for intravalley occupations read:
\begin{align}
    \left.\mathrm i\hbar\partial_t \Nextwo{\mu,\mathbf Q}{\xi,\xi,s}\right|_{\text{coul,Dex}} = 2\mathrm i\sum_{\nu}\mathrm{Im}\left(D_{\mu,\nu,\mathbf K^{\xi}-\mathbf K^{\bar \xi}}^{\xi,\xi,\bar \xi,\bar \xi,s,s}\Cextwo{\mu,\nu,\mathbf Q}{\xi,\bar \xi,s,s}\right),
\end{align}
the equations of motion for intervalley occupations read:
\begin{align}
    \left.\mathrm i\hbar\partial_t \Nextwo{\mu,\mathbf Q}{\xi,\bar \xi,s}\right|_{\text{coul,Dex}} = 2\mathrm i\sum_{\nu}\mathrm{Im}\mleft(D_{\mu,\nu,\mathbf K}^{\xi,\bar \xi,\Gamma, \xi,s,s}\Kextwo{\mu,\nu,\mathbf Q}{\xi,\bar \xi,\Gamma,\xi,s,s}\mright),
\end{align}
and:
\begin{align}
    \left.\mathrm i\hbar\partial_t \Nextwo{\mu,\mathbf Q}{\Gamma,\xi,s}\right|_{\text{coul,Dex}} = - 2\mathrm i\sum_{\nu}\mathrm{Im}\mleft(D_{\nu,\mu,\mathbf K}^{\xi,\bar \xi,\Gamma, \xi,s,s}\Kextwo{\nu,\mu,\mathbf Q}{\xi,\bar \xi,\Gamma,\xi,s,s}\mright).
\end{align}
The equations of motion for the correlations of intravalley excitons read:
\begin{align}
    \left.\mathrm i\hbar\partial_t \Cextwo{\mu,\nu,\mathbf Q}{\xi,\bar \xi,s,s}\right|_{\text{coul,Dex}} = D_{\nu,\mu,\mathbf K^{\xi}-\mathbf K^{\bar \xi}}^{\bar \xi,\bar \xi, \xi, \xi,s,s}\left(\Nextwo{\mu,\mathbf Q}{\xi,\xi,s} - \Nextwo{\nu,\mathbf Q}{\bar \xi,\bar \xi,s}\right),
\end{align}
and the equations of motion for the correlations of intervalley excitons read:
\begin{align}
    \left.\mathrm i\hbar\partial_t \Kextwo{\mu,\nu,\mathbf Q}{\xi,\bar \xi,\Gamma,\xi,s,s}\right|_{\text{coul,Dex}} = D_{\nu,\mu,\mathbf K}^{\Gamma,\xi,\xi,\bar \xi,s,s}\left(\Nextwo{\mu,\mathbf Q}{\xi,\bar \xi,s} - \Nextwo{\nu,\mathbf Q}{\Gamma,\xi,s}\right).
\end{align}
Since the Dexter matrix element does not vanish at $\mathbf Q=\mathbf 0$, cf.~Supplementary Figure~\ref{fig:ExchangeDexterMatrixElements}, an additional coherent contribution arises:
\begin{align}
    \left.\mathrm i\hbar\partial_t \Poltwo{\mu}{\xi,s}\right|_{\text{coul,Dex}} = \sum_{\nu}D_{\mu,\nu,\mathbf K^{\xi}-\mathbf K^{\xi^{\prime}}}^{\xi,\xi,\bar \xi,\bar \xi,s,s}\Poltwo{\nu}{\bar \xi,s}.
\end{align}

We note that all Coulomb scattering mechanisms (intra- and intervalley exchange, Dexter coupling) between excitonic occupations $\Nextwo{\mathbf Q}{j}$ and $\Nextwo{\mathbf Q}{i}$, cf.~Eq.~\eqref{eq:N_Def}, are mediated by two-exciton correlations $\Cextwo{\mathbf Q}{i,j}$ or $\Kextwo{\mathbf Q}{i,j}$, cf.~Eq.~\eqref{eq:C_Def} and Eq.~\eqref{eq:K_Def}, respectively, which annihilate an exciton in state $j$ and create an exciton in state $i$. These intraexcitonic transitions or coherences can also be viewed as virtual excitons \cite{haug2009quantum}.
Since these virtual excitons exhibit a finite lifetime due to the coupling to the phonon bath, cf.~Eq.~\eqref{eq:C_PhononEquations} and Eq.~\eqref{eq:K_PhononEquations}, which is far shorter than the actual lifetime of real excitonic occupations (which occurs due to, e.g.,\ Auger or radiative recombination within the light cone), the energy conservation is softened and off-resonant scattering between real excitonic occupations $\Nextwo{\mathbf Q}{j}$ and $\Nextwo{\mathbf Q}{i}$ becomes possible. 


\subsubsection{Comparing the Coulomb scattering matrix elements}

\begin{figure}[ht]
    \centering
    \includegraphics[width=0.47\linewidth]{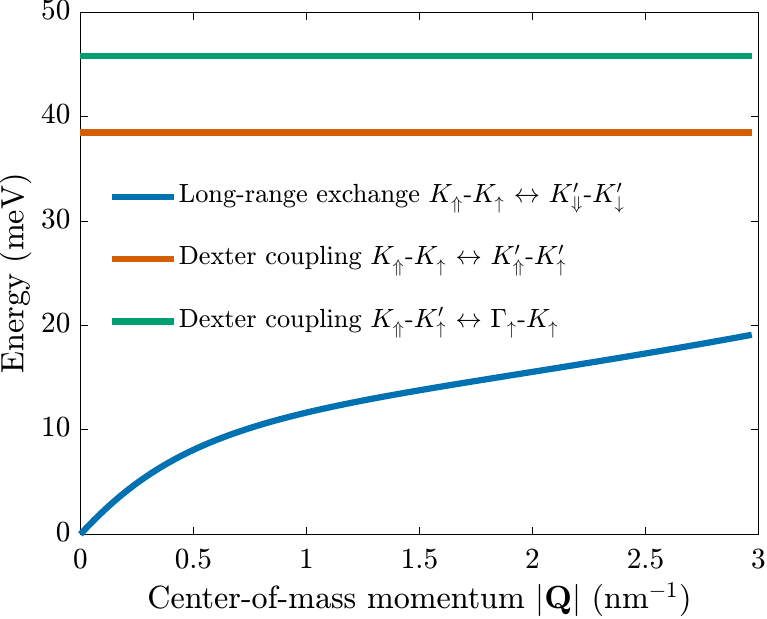}
    \caption{Intervalley exchange (blue line), Dexter coupling of intravalley excitons (red line) and Dexter coupling of intervalley excitons (green line) for WSe$_2$ on SiO$_2$.}
    \label{fig:ExchangeDexterMatrixElements}
\end{figure}

In Fig.~\ref{fig:ExchangeDexterMatrixElements} we depict the matrix elements for resonant intervalley exchange between intravalley A excitons (blue line) in Eq.~\eqref{eq:ExchangeMatrixElement}, and off-resonant Dexter coupling between A and B excitons (green line) in Eq.~\eqref{eq:DexterMatrixElement}. Here, intervalley exchange monotonously scales with the center-of-mass momentum $\mathbf Q$. Dexter coupling is almost independent of the center-of-mass momentum and, especially, does not vanish at \mbox{$\mathbf Q=\mathbf 0$}. The calculated magnitude of about 40\,meV is in line with other works \cite{berghauser2018,timmer2024ultrafast}. Even though the Dexter matrix element is larger than the exchange matrix element, the off-resonant nature of the Dexter coupling mechanism reduces its effectiveness significantly, so that Dexter interaction is only relevant with regard to the valley depolarization on a several-picosecond timescale (at 250\,K) as discussed in the main manuscript.

\FloatBarrier
\subsubsection{Macroscopically screened Coulomb potential}
\label{sec:ScreenedCoulombPotential}

The screened Coulomb potential governing all Coulomb interaction processes reads:
\begin{align}
    V_{\mathbf q} = \frac{e^2}{\mathcal A}\int\mathrm dz\,\mathrm dz^{\prime}\,\left|\xi(z)\right|^2G_{\mathbf q}\left(z,z^{\prime}\right)\left|\xi(z^{\prime})\right|^2,
    \label{eq:V_Screened}
\end{align}
where $\xi(z)$ is the carrier confinement wave function and $G_{\mathbf q}\left(z,z^{\prime}\right)$ is the Green's function of the corresponding Poisson equation.
To take the influence of the dielectric environment into account, we solve the Poisson equation for a five-layer geometry, i.e.,\ for a monolayer $\epsilon_3$ with thickness $d$ and interlayer separation $h$ to substrate $\epsilon_1$ and superstrate $\epsilon_5$. Here, we explicitly take a small vacuum gap $h$ between the 1L-TMD and the surrounding materials into account, which is crucial not to overestimate the substrate screening \cite{florian2018dielectric}. The corresponding Green's function reads:
\begin{align}
\begin{split}
    &G_{\mathbf q}(z,z^{\prime}) =  \frac{1}{2\epsilon_0\epsilon_3|\mathbf q|}\mathrm e^{-|\mathbf q||z-z^{\prime}|}\\
    &\, +\frac{1}{f_{\mathbf q}}\left(\left(\epsilon_{3,-}\epsilon_{5,+}\mathrm e^{h|\mathbf q|}-\epsilon_{3,+}\epsilon_{5,-}\mathrm e^{-h|\mathbf q|}\right)\left(\epsilon_{1,+}\epsilon_{3,-}\mathrm e^{h|\mathbf q|}-\epsilon_{1,-}\epsilon_{3,+}\mathrm e^{-h|\mathbf q|}\right)\mathrm e^{-d|\mathbf q|}\left(\mathrm e^{|\mathbf q|(z-z^{\prime})}+\mathrm e^{-|\mathbf q|(z-z^{\prime})}\right)\right.\\
    &\,\left.
    +\left(\epsilon_{3,-}\epsilon_{5,+}\mathrm e^{h|\mathbf q|}-\epsilon_{3,+}\epsilon_{5,-}\mathrm e^{-h|\mathbf q|}\right)\left(\epsilon_{1,+}\epsilon_{3,+}\mathrm e^{h|\mathbf q|}-\epsilon_{1,-}\epsilon_{3,-}\mathrm e^{-h|\mathbf q|}\right)\mathrm e^{|\mathbf q|(z+z^{\prime})}\right.\\
    &\,\left.
    +\left(\epsilon_{3,+}\epsilon_{5,+}\mathrm e^{h|\mathbf q|}-\epsilon_{3,-}\epsilon_{5,-}\mathrm e^{-h|\mathbf q|}\right)\left(\epsilon_{1,+}\epsilon_{3,-}\mathrm e^{h|\mathbf q|}-\epsilon_{1,-}\epsilon_{3,+}\mathrm e^{-h|\mathbf q|}\right)\mathrm e^{-|\mathbf q|(z+z^{\prime})}
    \right),
    \end{split}
\end{align}
with
\begin{multline}
    f_{\mathbf q} =  2\epsilon_0\epsilon_3|\mathbf q|\left(-\epsilon_{3,-}\epsilon_{3,-}\left(\epsilon_{1,+}\epsilon_{5,+}\mathrm e^{-d|\mathbf q|}\mathrm e^{2h|\mathbf q|} - \epsilon_{1,-}\epsilon_{5,-}\mathrm e^{d|\mathbf q|}\mathrm e^{-2h|\mathbf q|}\right)\right.\\
    -2\epsilon_{3,+}\epsilon_{3,-}\left(\epsilon_1\epsilon_5-1\right)\left(\mathrm e^{d|\mathbf q|}-\mathrm e^{-d|\mathbf q|}\right)
    \left. 
    + 
    \epsilon_{3,+}\epsilon_{3,+}\left(\epsilon_{1,+}\epsilon_{5,+}\mathrm e^{d|\mathbf q|}\mathrm e^{2h|\mathbf q|} - \epsilon_{1,-}\epsilon_{5,-}\mathrm e^{-d|\mathbf q|}\mathrm e^{-2h|\mathbf q|}\right)
    \right),
\end{multline}
where
\begin{align}
    \epsilon_{i,\pm}=\epsilon_i\pm 1.
\end{align}
For $\cos$-confinement with $\xi(z) = \sqrt{\dfrac{2}{d}}\cos\left(\dfrac{\pi}{d}z\right)$, which models the charge distributions in $z$-direction resembling DFT results \cite{latini2015excitons}, the integrals read:
\begin{align}
\begin{split}
    &\frac{4}{d^2}\int_{-\frac{d}{2}}^{\frac{d}{2}}\mathrm dz\,\mathrm dz^{\prime}\,\cos^2\left(\frac{\pi}{d}z\right)\mathrm e^{-|\mathbf q||z-z^{\prime}|}\cos^2\left(\frac{\pi}{d}z^{\prime}\right) =
    \frac{1}{|\mathbf q|d\left(4\pi^2+|\mathbf q|^2d^2\right)}\left(8\pi^2+3|\mathbf q|^2d^2 - \frac{32\pi^4\left(1-\mathrm e^{-|\mathbf q|d}\right)}{|\mathbf q|d\left(4\pi^2+|\mathbf q|^2d^2\right)}\right),
    \end{split}
\end{align}
and
\begin{align}
    \frac{4}{d^2}\int_{-\frac{d}{2}}^{\frac{d}{2}}\mathrm dz\,\mathrm dz^{\prime}\,\cos^2\left(\frac{\pi}{d}z\right)\mathrm e^{\pm|\mathbf q|(\pm z\pm z^{\prime})}\cos^2\left(\frac{\pi}{d}z^{\prime}\right) = \left(\frac{8\pi^2\sinh\left(\frac{|\mathbf q|d}{2}\right)}{|\mathbf q|d\left(4\pi^2+|\mathbf q|^2d^2\right)}\right)^2.
\end{align}
With these ingredients, an explicit expression for the screened Coulomb potential in Eq.~\eqref{eq:V_Screened} is derived.
To model the screening at higher momenta, which is crucial for obtaining realistic exciton and biexciton binding energies, we use the model dielectric function from Ref.~\cite{trolle2017model}, i.e.,~the dielectric constant of the 1L-TMD gains a momentum-dependence $\epsilon_3\rightarrow\epsilon_{3,\mathbf q}$:
\begin{align}
\label{eq:dielectric_function_material_q_dependent}
    \epsilon_{3,\mathbf q} = 1 + \dfrac{1}{\left(\epsilon_{3}-1\right)^{-1} + \alpha_{\text{TF}}\dfrac{|\mathbf q|^2}{q_{\text{TF}}^2} + \dfrac{\hbar^2|\mathbf q|^4}{4m_0^2\omega_{\text{pl}}^2}}.
\end{align}
Here, $\epsilon_{3} = \epsilon_{3,\mathbf q=\mathbf 0}$ is the dielectric constant of the bulk material, $q_{\text{TF}} = \sqrt{\dfrac{3^{\frac{1}{3}}\omega_{\text{pl}}^{\frac{2}{3}}m_0^{\frac{4}{3}}}{\hbar^2\pi^{\frac{4}{3}}\epsilon_0^{\frac{2}{3}}}}$ is the Thomas-Fermi wave vector, $\omega_{\text{pl}}$ is the bulk plasmon frequency and $m_0$ is the free electron mass. 
$\alpha_{\text{TF}}$ is the fitting parameter, which is set to best reproduce the effective screening $\epsilon_{\text{eff},\mathbf q}$, defined by:
\begin{align}
    \epsilon_{\text{eff},\mathbf q} = \frac{V_{0,\mathbf q}}{V_{\mathbf q}},
\label{eq:EffScreen}
\end{align}
from \textit{ab initio} calculations extracted from the Computational Materials Repository (CMR) \cite{andersen2015dielectric}.
In Eq.~\eqref{eq:EffScreen}, $V_{0,\mathbf q}$ is the unscreened confined potential, which is obtained from $V_{\mathbf q}$ in Eq.~\eqref{eq:V_Screened} by setting $\epsilon_i=1$ for all $i$.


\FloatBarrier
\subsubsection{Details of the numerical simulations}

We solve the Wannier equation on a polar momentum grid with 800 radii and angles up to a relative momentum value $\mathbf q$ of 6~nm$^{-1}$. The exciton-phonon matrix elements are calculated for 200~radii and angles up to a relative momentum value of 6~nm$^{-1}$. The exciton-phonon scattering rates and Coulomb coupling elements, which enter the equations of motion, are calculated for 100~radii and angles for a center-of-mass momentum $\mathbf Q$ up to 3~nm$^{-1}$. The coupled equations of motion are solved via a Runge-Kutta algorithm of fourth order with a constant time step size of 1~fs.
We note that in contrast to Refs.~\cite{selig2019ultrafast,selig2020}, where the intervalley exchange mechanism is treated in an adiabatic approximation, we keep the equations of motion of the two-exciton correlations $C$ and $K$ as they are. Regarding the intra-/intervalley exchange, which is in principle angle-dependent, it is sufficient to consider an angle of zero, which massively reduces the computational effort. We also note, that keeping the equations of motion of the two-exciton correlations $C$ and $K$ in their original form (especially regarding resonant intervalley exchange) yields a more stable Runge-Kutta routine compared to their adiabatic elimination, while, at the same time, being physically more accurate.


\subsubsection{Occupation dynamics}

In supplementary Figure~\ref{fig:supp_occupation_dynamics77K} and supplementary Figure~\ref{fig:supp_occupation_dynamics250K}, we depict the total valley- and spin-resolved excitonic densities $N^{\xi,\xi^{\prime},s}$, which are given by:
\begin{align}
    N^{\xi,\xi^{\prime},s} = \frac{1}{\mathcal A}\sum_{\mathbf Q}\left(\big|\Poltwo{}{\xi,s}\big|^2\delta_{\xi,\xi^{\prime}}\delta_{\mathbf Q,\mathbf 0}+\Nextwo{\mathbf Q}{\xi,\xi^{\prime},s}\right),
    \label{eq:total_excitonic_density}
\end{align}
for a $\sigma^+$-polarized optical pump pulse at a temperature of 77~K and 250~K, respectively. Note, that the excitonic densities involving electrons at the $\Lambda$/$\Lambda^{\prime}$ valleys read: $N^{\xi,\Lambda/\Lambda^{\prime},s} = 3\frac{1}{\mathcal A}\sum_{\mathbf Q}\Nextwo{\mathbf Q}{\xi,\Lambda/\Lambda^{\prime},s}$, since there are three equivalent $\Lambda/\Lambda^{\prime}$ valleys in the first Brillouin zone.

In supplementary Figure~\ref{fig:supp_occupation_dynamics77K}, after the optical excitation of intravalley $K_{\Uparrow}^{}$-$K_{\uparrow}^{}$ coherent occupations and a phonon-assisted dephasing into incoherent $K_{\Uparrow}$-$K_{\uparrow}$ occupations, intervalley electron scattering into $K_{\Uparrow}^{}$-$K^{\prime}_{\uparrow}$ and $K_{\Uparrow}^{}$-$\Lambda^{}_{\uparrow}$ occupations within the first 100--200~fs occurs. 
Then, on a picosecond scale, a valley equilibration via the combined action of momentum-dark Dexter scattering, cf.~supplementary Figure~\ref{fig:supp_depolarization}(b), phonon-assisted hole scattering, cf.~Supplementary Figure~\ref{fig:supp_depolarization}(c), and intervalley exchange interaction, cf.~Supplementary Figure~\ref{fig:supp_depolarization}(d), as described in the main manuscript and in Supplementary Section~\ref{sec:discussion_valley_depolarization}, occurs.

Comparing the 77~K-case in supplementary Figure~\ref{fig:supp_occupation_dynamics77K} to the 250~K-case in supplementary Figure~\ref{fig:supp_occupation_dynamics250K}, the overall initial intravalley occupation at the $K$ valley is larger at 77~K (upper left). To quantify this observation, we define a degree of valley polarization $\eta$ at a time of zero within an optical excitation of a $\sigma^+$ pulse by:
\begin{align}
    \eta = \frac{N^{K,K,\Uparrow,\uparrow}(t=0)-N^{K^{\prime},K^{\prime},\Downarrow,\downarrow}(t=0)}{\sum_{\xi,\xi^{\prime},s}N^{\xi,\xi^{\prime},s}(t=0)}.
    \label{eq:valley_polarization}
\end{align}
Hence, within this definition, valley polarization is 1 ($-1$), if all excitons reside at the $K$ ($K^{\prime}$) valley in spin-up (spin-down) configuration, and zero, if the intravalley excitonic occupations at the $K$ and at the $K^{\prime}$ valleys are equal or if all excitons form intervalley occupations. 
Note, that the definition of valley polarization in Eq.~\eqref{eq:valley_polarization} is not equal to the degree of valley polarization defined by the circular dichroism as discussed in the main manuscript, since the circular dichroism is sensible to the bleaching of individual electrons and holes as constituents of the corresponding exciton.


Within our calculations, we obtain: \mbox{$\eta = 0.80~(\text{5~K}),~0.77~(\text{77~K}),~0.58~(\text{250~K})$}, i.e., the overall initial degree of valley polarization is never 100~\% and decreases with increasing temperature. The reason for this behavior is exciton-phonon interaction, as efficient intervalley electron scattering into $\Lambda$ and $K^{\prime}$ valleys already depletes some fraction of the optically pumped intravalley occupation even at time zero (pulse envelope maximum). At increasing temperature, the exciton distribution broadens with respect to the center-of-mass momentum $\mathbf Q$ due to thermal activation via acoustic phonons. This in turn increases the strength of intervalley scattering into $K$-$\Lambda$ excitons, as a larger fraction of occupied excitonic states at larger center-of-mass momenta renders intervalley scattering into $K$-$\Lambda$ excitons energetically more favorable.

The decrease of the initial valley polarization in Eq.~\eqref{eq:valley_polarization} with increasing temperature is in agreement with other measurements \cite{lagarde2014carrier, mujeeb2023influence} and first-principle calculations \cite{lin2022phonon}.


\begin{figure}[ht]
    \centering
    \includegraphics[width = 0.45\linewidth]{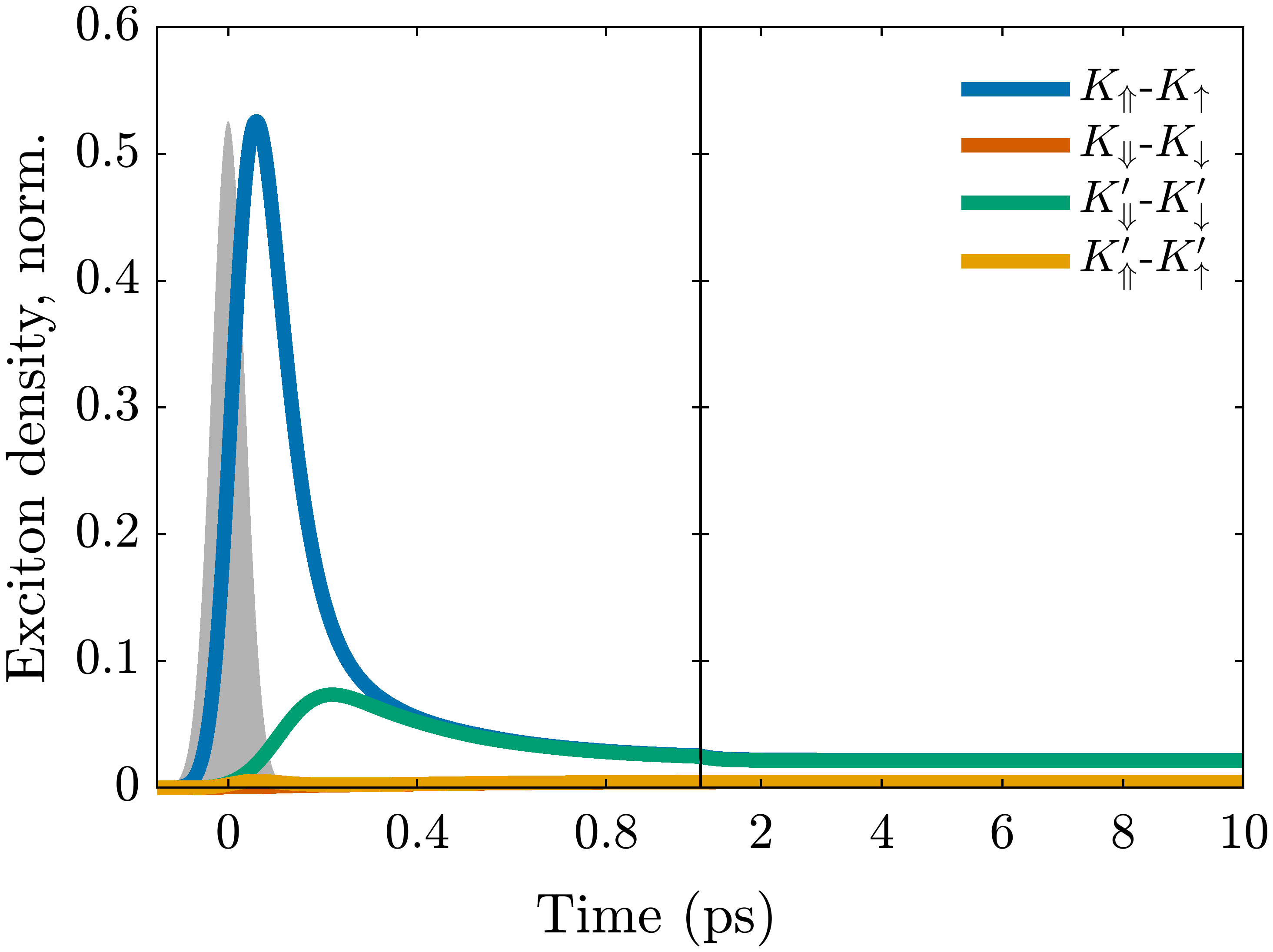}
    \includegraphics[width = 0.45\linewidth]{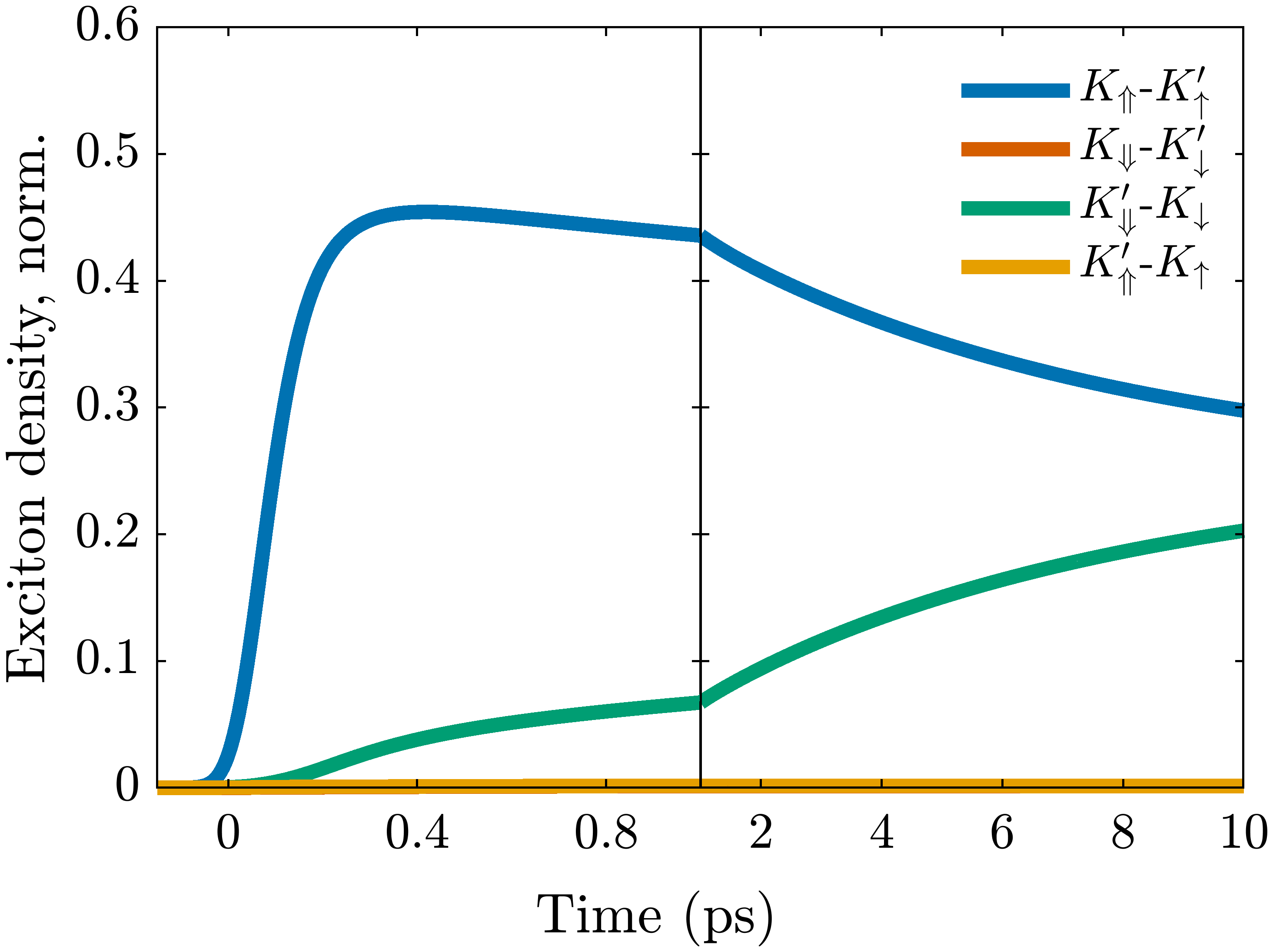}
    \includegraphics[width = 0.45\linewidth]{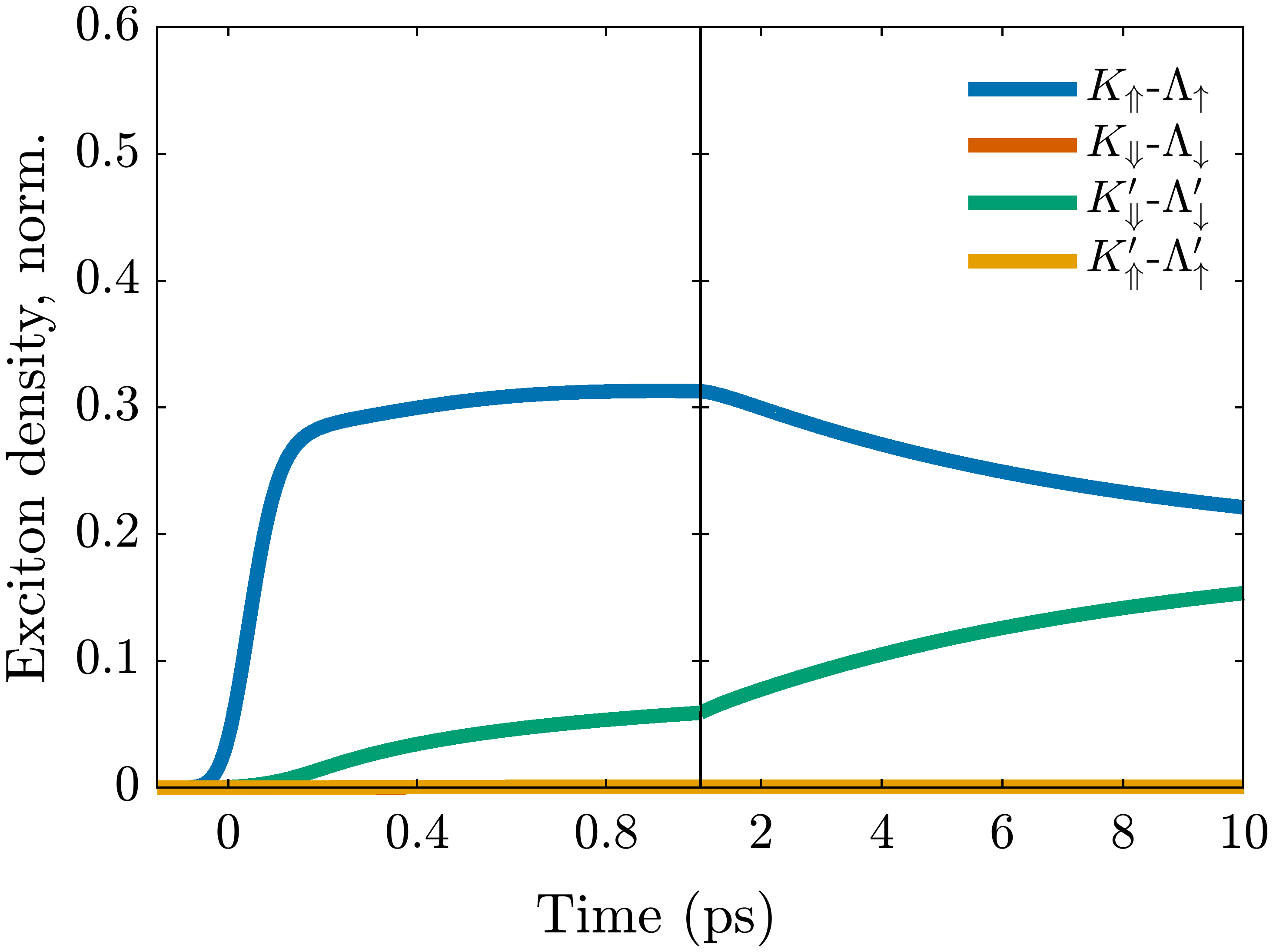}
    \includegraphics[width = 0.45\linewidth]{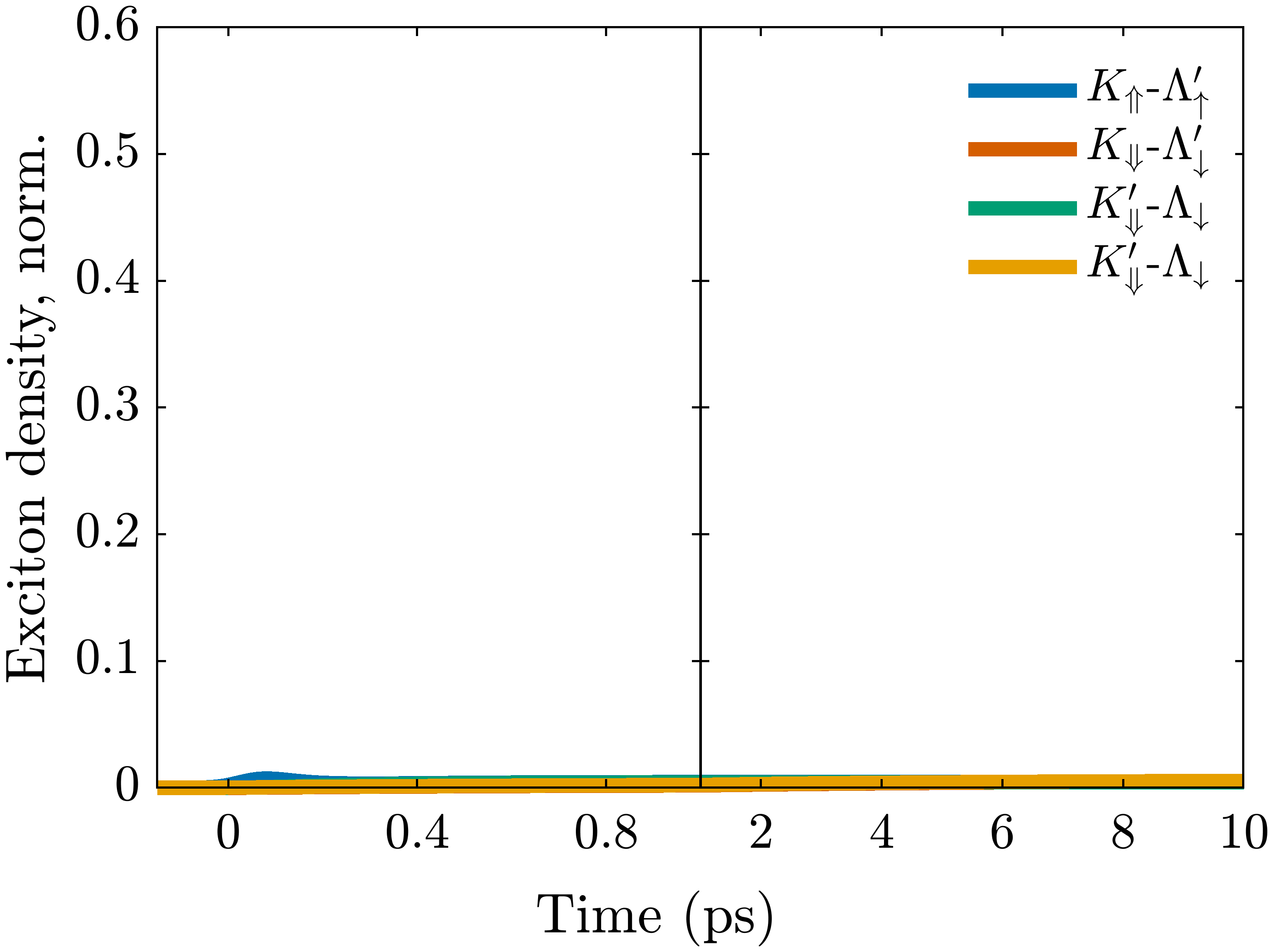}
    \includegraphics[width = 0.45\linewidth]{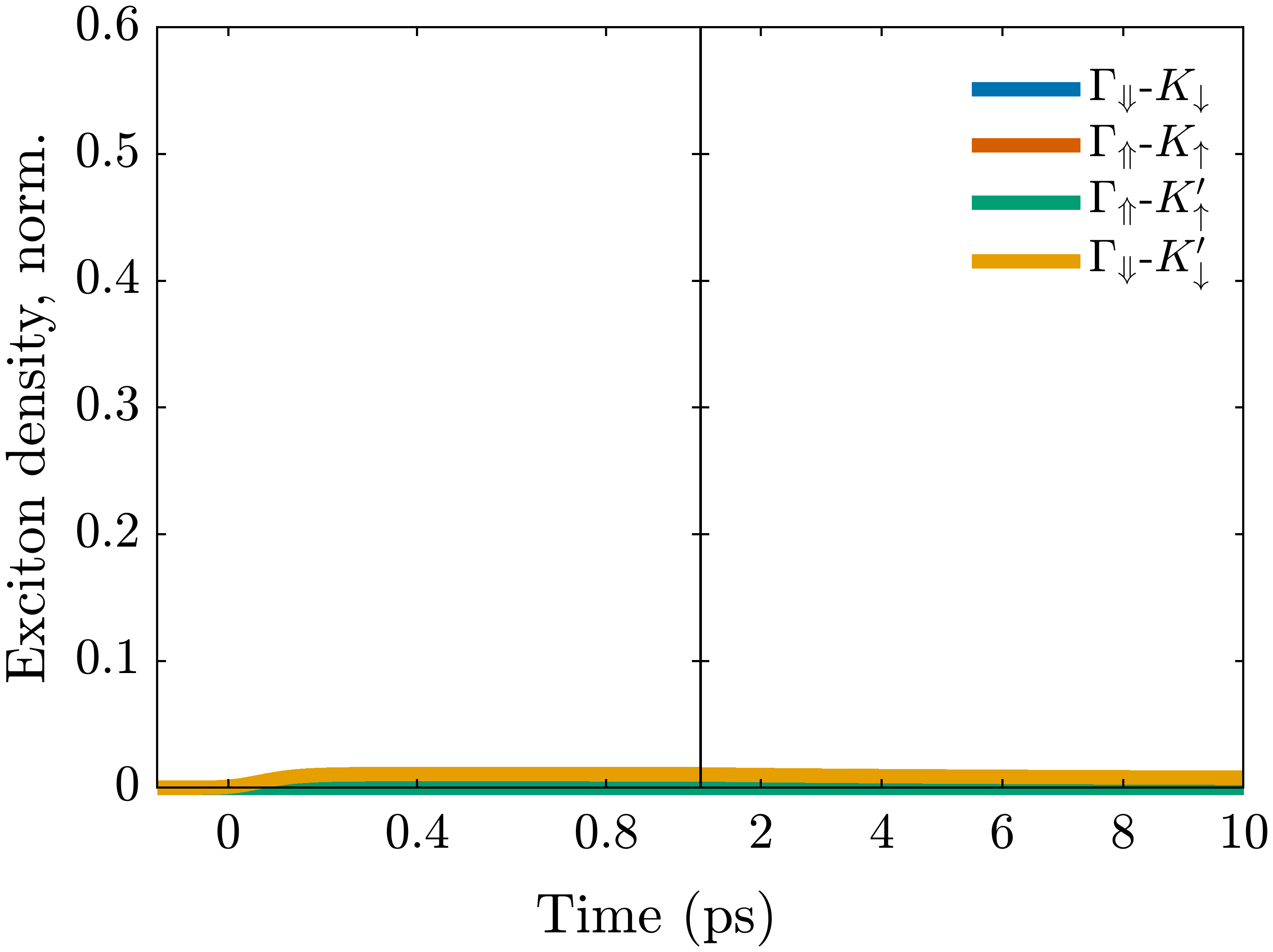}
    \caption{Dynamics of the total spin- and valley-resolved excitonic densities $N^{\xi,\xi^{\prime},s,s^{\prime}}$, cf.~Eq.~\eqref{eq:total_excitonic_density}, for a $\sigma^+$-polarized optical pump pulse (grey-shaded area) at the A exciton at a temperature of 77\,K.}
    \label{fig:supp_occupation_dynamics77K}
\end{figure}

\begin{figure}[ht]
    \centering
    \includegraphics[width = 0.45\linewidth]{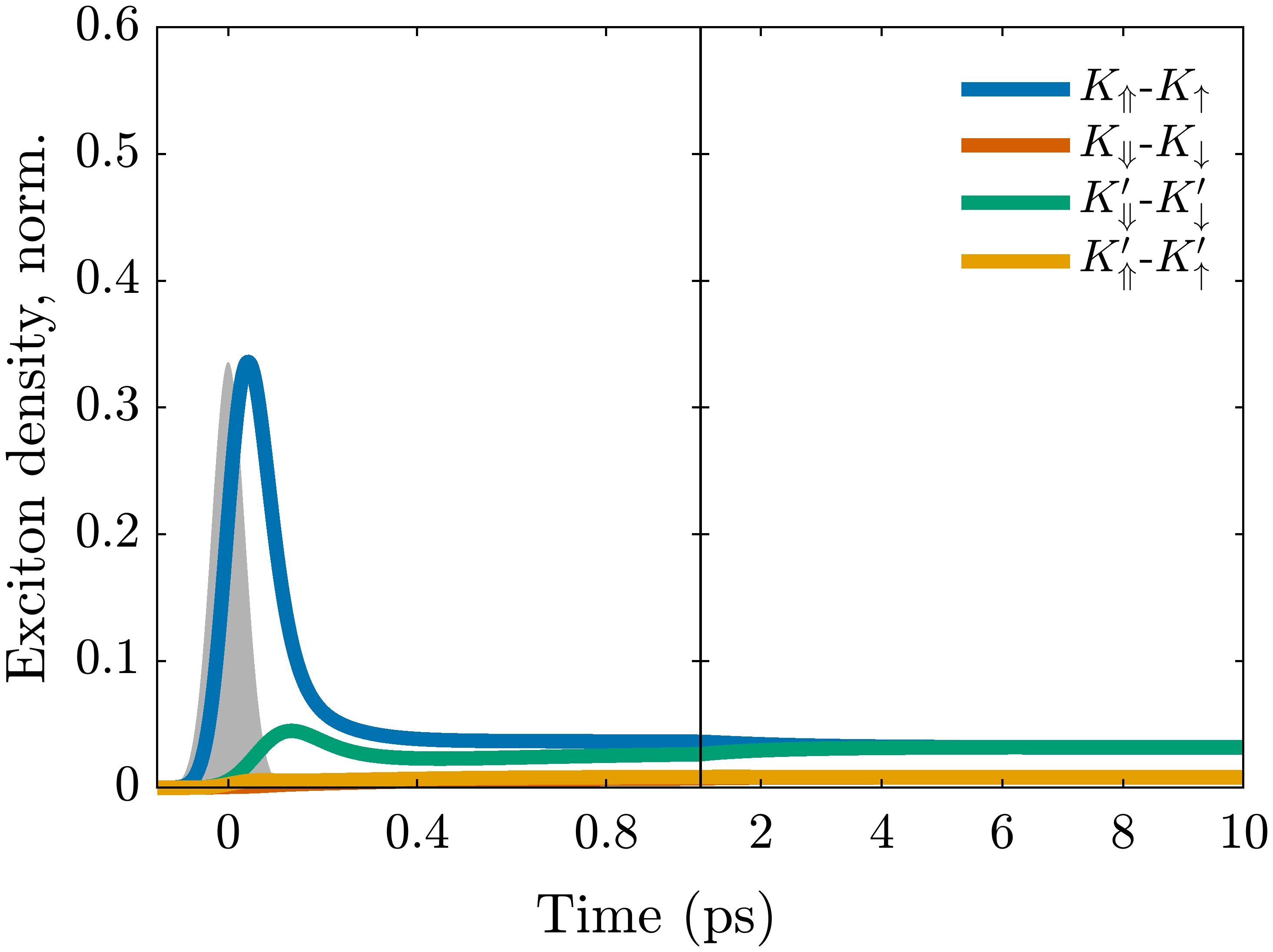}
    \includegraphics[width = 0.45\linewidth]{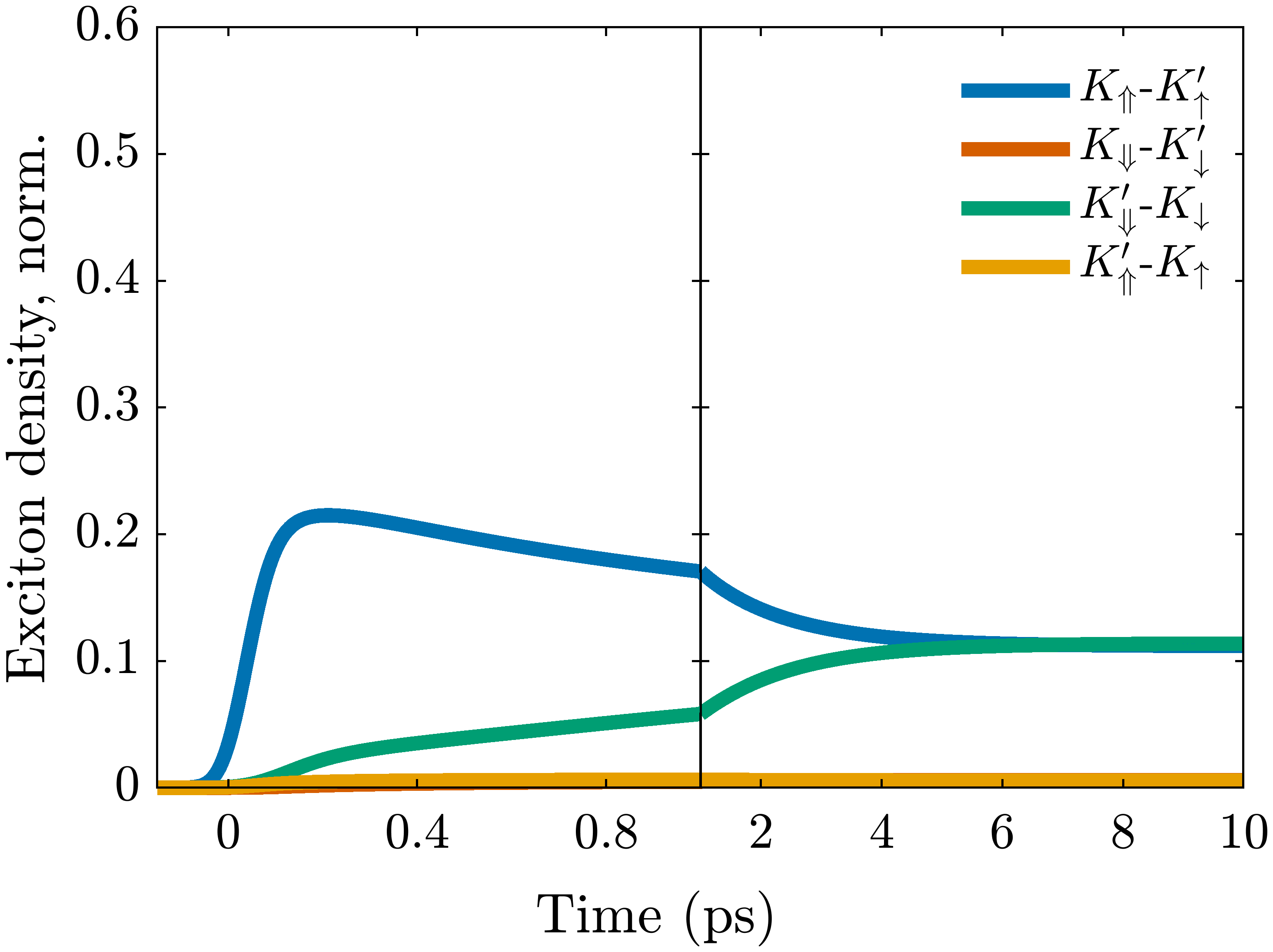}
    \includegraphics[width = 0.45\linewidth]{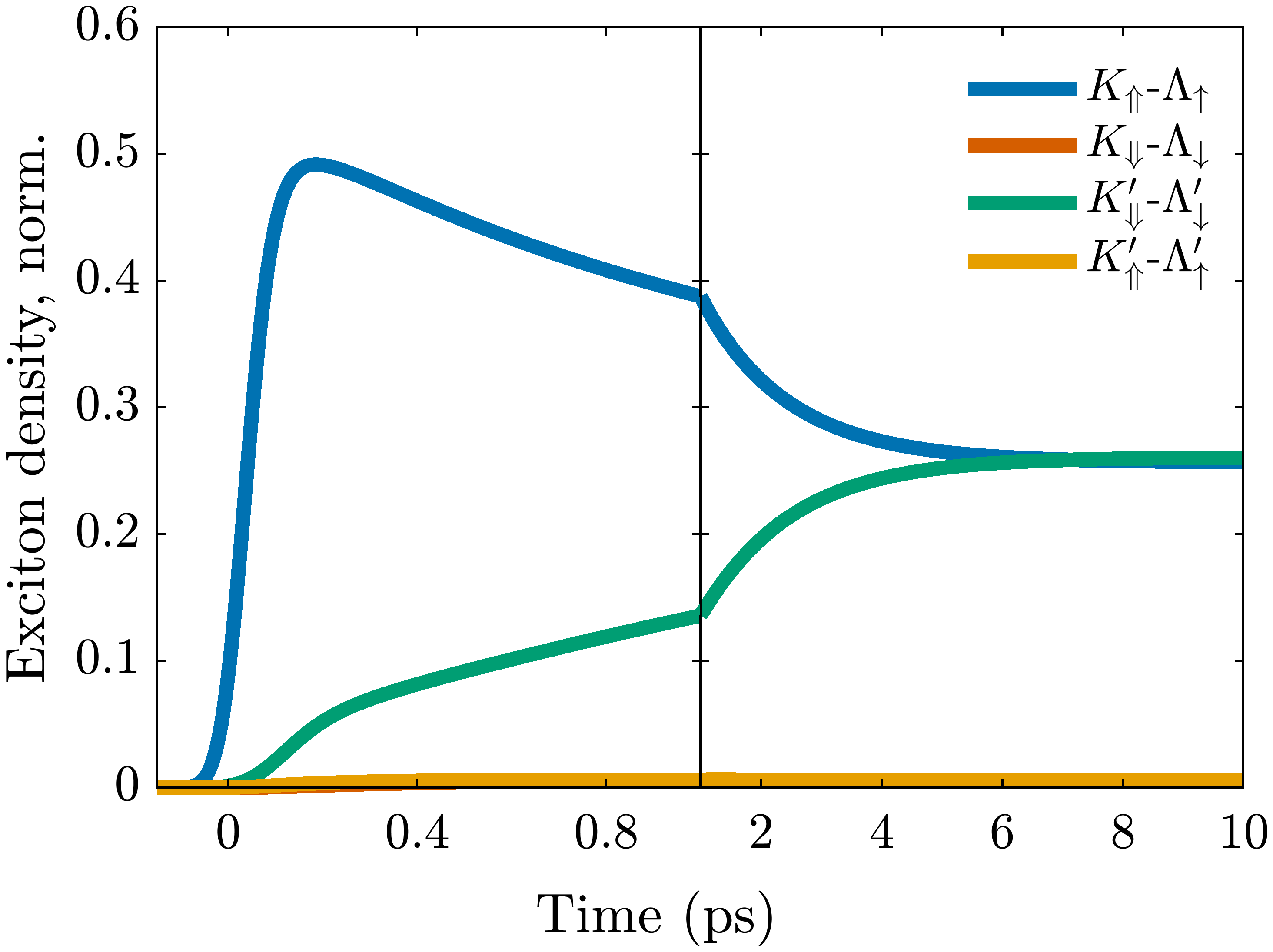}
    \includegraphics[width = 0.45\linewidth]{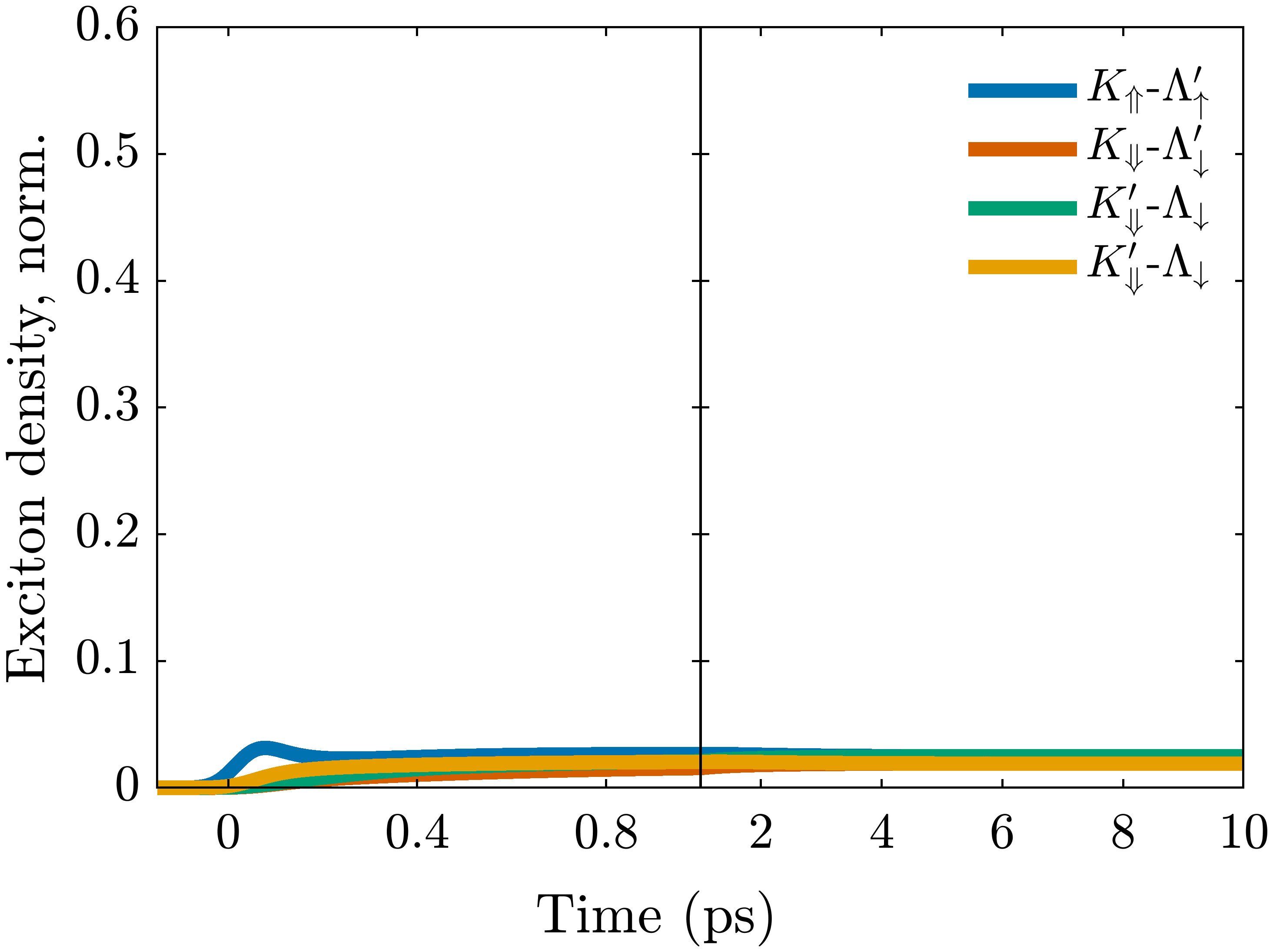}
    \includegraphics[width = 0.45\linewidth]{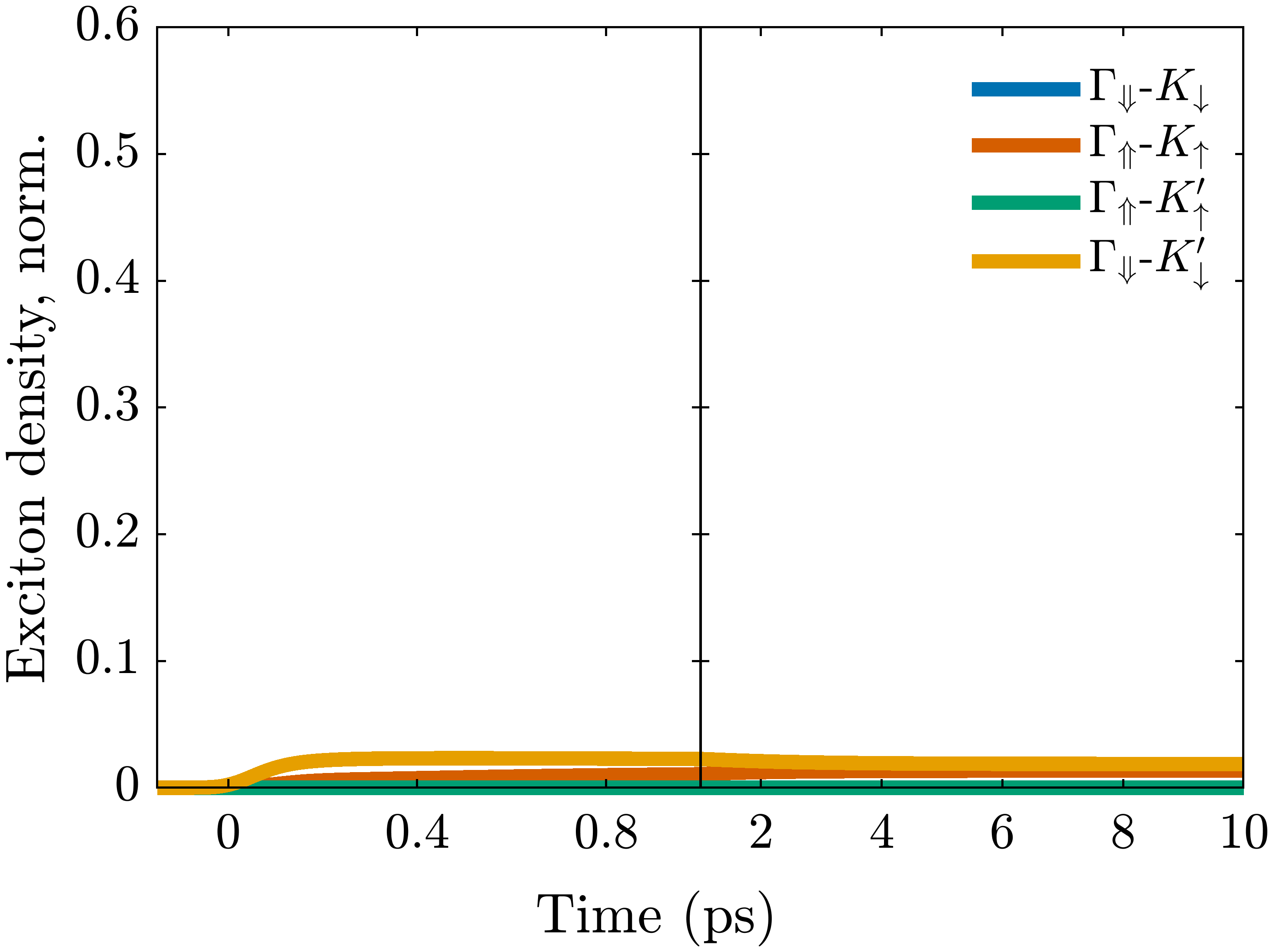}
    \caption{Dynamics of the total spin- and valley-resolved excitonic densities $N^{\xi,\xi^{\prime},s,s^{\prime}}$, cf.~Eq.~\eqref{eq:total_excitonic_density}, for a $\sigma^+$-polarized optical pump pulse (grey-shaded area) at the A exciton at a temperature of 250\,K.}
    \label{fig:supp_occupation_dynamics250K}
\end{figure}


\subsubsection{Parameters}

We use the effective masses, band splittings and momentum matrix elements from Ref.~\cite{kormanyos2015k}. All phonon-scattering-related parameters such as phonon dispersions and electron-phonon interaction potentials in deformation potential approximation are taken from Ref.~\cite{jin2014intrinsic}. All other parameters used in the numerical simulations are shown in Table~\ref{tab:parameters}.
\begin{table}[h!]
    \centering
    \begin{tabular}{l l l l}
    Pump pulse duration & $\sigma_{\,\text{P}}$ & 50$\,$fs & (exp.)\\
    Static dielectric constant of WSe$_2$ & $\epsilon_{3,\mathbf q=\mathbf 0}$ & $\sqrt{7.8\cdot 15.9}$ & \cite{laturia2018dielectric}\\
    Plasma frequency of WSe$_2$ & $\hbar\omega_{\text{pl}}$ & 22.6$\,$eV & \cite{kumar2012tunable}\\
    Monolayer width & $d$ & 0.6575$\,$nm & \cite{kylanpaa2015binding}\\
    Distance to SiO$_2$ substrate & $h$ & 0.3$\,$nm & \cite{florian2018dielectric,druppel2017diversity}\\
    Thomas-Fermi screening fitting parameter & $\alpha_{\text{TF}}$ & 1.9 & (fit to CMR \cite{andersen2015dielectric})\\
    Static dielectric constant of SiO$_2$ & $\epsilon_{\text{SiO$_2$},0}$ & 3.9 & \cite{xue2011scanning}\\
    Optical dielectric constant of SiO$_2$ & $\epsilon_{\text{SiO$_2$},\infty}$ & 2.12 & \cite{polyanskiy2024refractiveindex}\\
    Conduction band spin splitting at the $K$-valley & & 13\,meV  & \cite{ren2023measurement,kapuscinski2021rydberg}
    \end{tabular}
    \caption{Additional parameters used in the numerical calculations.}
    \label{tab:parameters}
\end{table}


\FloatBarrier
\section{Discussion of the valley depolarization}
\label{sec:discussion_valley_depolarization}
This section extends the discussion of the valley depolarization dynamics, presented in the main text.

The first process that describes valley depolarization is a non-local intervalley exchange interaction, in which the A exciton intravalley occupation at the $K$ valley scatters into an A exciton occupation at the $K'$ valley via double spin flips, as shown in Figure~4c, main text (see also Supplementary Note~I-B4).
However, since intravalley occupations generated upon an optical excitation are quickly scattered into intervalley occupations via phonon-assisted intervalley scattering (first process in Supplementary Figure~\ref{fig:supp_depolarization}), intervalley exchange is strongly suppressed and not sufficient to explain the overall decay of the A signal (see also Ref.~\cite{selig2020}). 

\begin{figure}[ht]
    \centering
    \includegraphics[width = 1.0\linewidth]{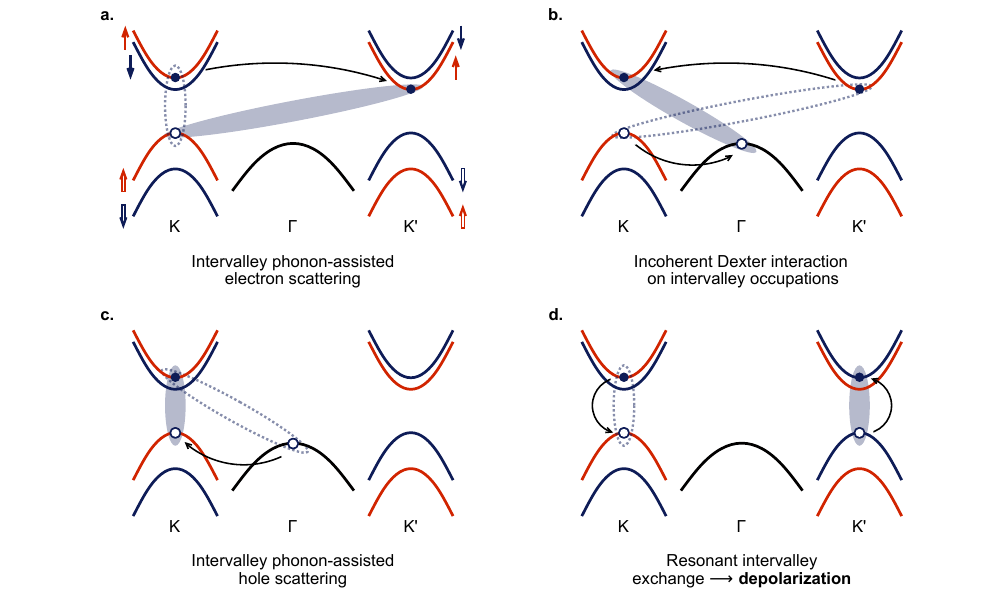}
    \caption{
    \textbf{Processes leading to valley depolarization.}
    The valley depolarization can be achieved by exchange interaction acting on intermediate intravalley occupations, generated via a combined action of the incoherent Dexter exchange and the phonon-assisted hole transfer.
    See details in text.
    }
    \label{fig:supp_depolarization}
\end{figure}

Alternatively, intervalley \mbox{$K^{}_\Uparrow$-$K^{\prime}_\uparrow$} (\mbox{$K^{\prime}_\Downarrow$-$K^{}_\downarrow$}) occupations can be scattered into intervalley \mbox{$\Gamma^{}_\Uparrow$-$K^{}_\uparrow$} (\mbox{$\Gamma^{}_\Downarrow$-$K^{\prime}_\downarrow$}) excitonic occupations via spin-conserved intraband Coulomb scattering of electrons and holes with a large intervalley momentum transfer (as in main text, in our notation, the first (second) index denotes the valley spin of the hole (electron) of the corresponding exciton).
Such process, which couples excitons of different valleys, is shown in Supplementary Figure~\ref{fig:supp_depolarization}b (incoherent Dexter coupling of intervalley occupations).
Although incoherent Dexter coupling of intervalley occupations cannot lead to a valley depolarization on its own, since the spins are conserved, in conjunction with other scattering processes, Dexter coupling speeds up the valley depolarization.
It can be understood as follows (see Supplementary Figure~\ref{fig:supp_depolarization}): 
Intervalley \mbox{$K^{}_\Uparrow$-$K^{\prime}_\uparrow$} (\mbox{$K^{\prime}_\Downarrow$-$K^{}_\downarrow$}) occupations, generated upon the phonon-assisted electron scattering (panel \textbf{a}), are upscattered into intervalley \mbox{$\Gamma^{}_\Uparrow$-$K^{}_\uparrow$} (\mbox{$\Gamma^{}_\Downarrow$-$K^{\prime}_\downarrow$}) occupations via Dexter coupling (panel \textbf{b}), which are subsequently scattered back into intravalley \mbox{$K^{}_\Uparrow$-$K^{}_\uparrow$} (\mbox{$K^{\prime}_\Downarrow$-$K^{\prime}_\downarrow$}) occupations via efficient phonon-assisted hole scattering between the $\Gamma$ and $K$/$K^{\prime}$ valleys (panel \textbf{c}).
Thus, due to the constant upscattering via Dexter coupling in conjunction with phonon-assisted hole scattering, there are always intermediate intravalley \mbox{$K^{}_\Uparrow$-$K^{}_\uparrow$} (\mbox{$K^{\prime}_\Downarrow$-$K^{\prime}_\downarrow$}) occupations, on which intervalley exchange (panel \textbf{d}), can act upon. 
Therefore, Dexter coupling retrieves the otherwise strongly suppressed intervalley exchange to some degree and thus enables an efficient valley depolarization.

We have also examined intravalley exchange \cite{guo2019exchange} between intravalley A \mbox{$K^{}_\Uparrow$-$K^{}_\uparrow$} (\mbox{$K^{\prime}_\Downarrow$-$K^{\prime}_\downarrow$}) and B \mbox{$K^{}_\Downarrow$-$K^{}_\downarrow$} (\mbox{$K^{\prime}_\Uparrow$-$K^{\prime}_\uparrow$}) excitonic occupations, but found no significant impact on the CD dynamics. 
Although intravalley exchange, similar to Dexter interaction, leads to a constant upscattering of a fraction of A excitonic occupations into B occupations in the same valley, its impact is almost non-detectable due to the short lifetime of the A exciton occupation and its non-resonant nature. 
In this case, intervalley Dexter interaction (Supplementary Figure~\ref{fig:supp_depolarization}b), even though it is also non-resonant, is much more efficient, since it acts on the largely populated energetically lowest intervalley occupations.

Summarizing the discussed processes, the depolarization dynamics, as shown in Figure~5, main text, cannot be explained solely by phonon assisted intervalley scattering processes. 
Instead, the joint action of the incoherent Dexter coupling, the intervalley exchange, and phonon-assisted hole scattering define the overall decay of the A exciton CD signal. 
Taking into account the combined effect of all these mechanisms, we are able to correctly simulate the decay time of the A exciton CD signal for different temperatures, as shown in Figure~5, main text. 
Since the spin-unlike excitons in 1L-TMDs generally have lower energy compared to the the optically addressable spin-like exciton \cite{feierabend2021}, we also expect the discussed processes to be play an important role in other 1L-TMDs. 


\section{Additional results}

\subsection{Time resolved Faraday rotation}

\begin{figure}[ht]
    \centering
    \includegraphics[width = 1.0\linewidth]{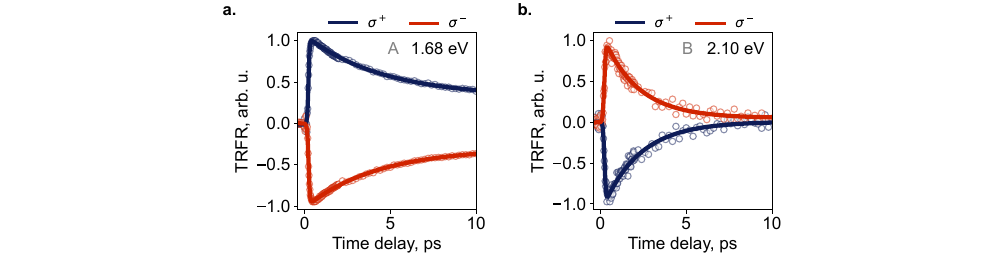}
    \caption{
    \textbf{TRFR results.}
    TRFR signal of A (\textbf{a}) and B (\textbf{b}) excitons, respectively, showing opposite signal sign for both excitation polarizations, 6~$\upmu\mathrm{J \, cm^{-2}}$ pump fluence, 77~K sample temperature; probe photon energy is indicated in top right corner of each panel. 
    Dots correspond to experimental points, solid lines show the fit.
    }
    \label{fig:faraday}
\end{figure}

To confirm the key observations of the time resolved CD measurements, reported in the main text, we perform TRFR experiments. 
Figure~\ref{fig:faraday} shows TRFR traces of A and B excitons, respectively, induced by $\sigma^+$ and $\sigma^-$ pump (6~$\upmu\mathrm{J \, cm^{-2}}$ fluence). 
The signals obtained upon excitation with the light of the same helicity is presented with the same color coding.
A sign flip of the TRFR signals of A and B excitons upon switching the pump helicity indicates that the signal of both excitons is related to the pump-induced valley polarization.
For both pump light helicities, the TRFR signals of A and B excitons have opposite signs, which is consistent with the CD measurements discussed above. 
In the TRFR measurements, the fast component of both signals is not observed, which can be explained by the fact that the temporal resolution of this technique is lower than that of the CD measurements, as well as the choice of the probe photon energy \cite{yan2017exciton, su2018gamma}.
The data is therefore approximated with a simplified Equation~(2), containing only one exponential function. 
The depolarization time of the A exciton signal ($4.5 \pm 0.3$~ps) is markedly longer than that of the B exciton ($2.2 \pm 0.2$~ps), similarly to the CD results (see Figure~2f,h in the main text).
The TRFR provides therefore an additional confirmation of some of the main observations reported in the manuscript. 

\subsection{B exciton signal temperature dependence}

\begin{figure}[ht]
    \centering
    \includegraphics[width = 1.0\linewidth]{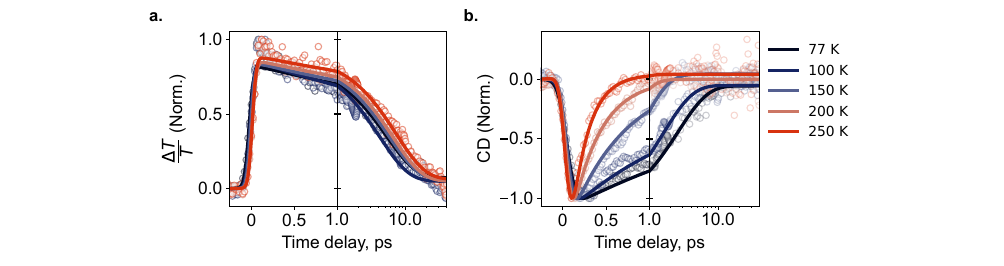}
    \caption{
    \textbf{B exciton temperature dependence.}
    \textbf{a} Normalized differential transmission traces, taken at the maximum of the transient signal. 
    \textbf{b} Normalized CD traces, calculated as explained in the main text.
    Dots represent experimental data points, solid lines show an approximation of the data with a monoexponential decay function.
    }
    \label{fig:tdep_bleach_B}
\end{figure}

Supplementary Figure~\ref{fig:tdep_bleach_B} compares the temperature dependence of the B exciton differential transmission signal measured for the $\sigma^+\sigma^+$ pump-probe configuration with the B exciton CD.
The data presented in panel (\textbf{a}) is taken from the $\Delta T / T$ maps used to calculate the CD signal, cf. Equation~(1), main text.
Panel (\textbf{b}) reproduces the data shown in Figure~3c, main text. 
The comparison of the two plots show that although the B exciton $\Delta T / T$ signal does not show a significant temperature dependence, the CD decay dynamics becomes much faster as the temperature increases, unambiguously indicating that the observed trend is due to the valley depolarization dynamics, rather than due to the overall acceleration of the carrier recombination.


\subsection{Simulated CD dynamics at high temperatures}

\begin{figure}[ht]
    \centering
    \includegraphics[width=.46\textwidth]{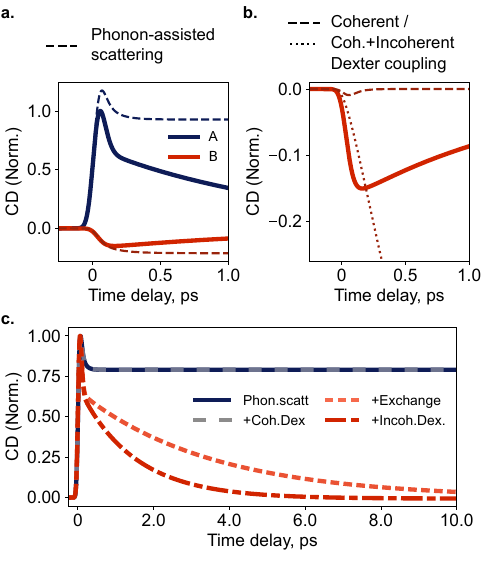}
    \caption{
    \textbf{Contribution of select processes to the CD signal at 250~K.}
    A comparison of total simulated CD traces with the individual contributions of the phonon-assisted intervalley scattering (\textbf{a}) and Dexter coupling (coherent and incoherent) (\textbf{b}).
    For the latter, intravalley phonon-assisted thermalization is also considered.
    All traces are normalized to the largest CD value. 
    \textbf{c} Contributions of phonon-assisted scattering, coherent Dexter coupling, intervalley exchange, Dexter intervalley exchange, and incoherent Dexter to the overall decay of the A exciton CD signal.
    }
    \label{fig:theory_components_250K}
\end{figure}

Supplementary Figure~\ref{fig:theory_components_250K} shows the contributions of the discussed intervalley coupling processes to the dynamics of the A and B CD signals at a temperature of 250~K. 
Similarly to Figure~6, main text, Supplementary Figure~\ref{fig:theory_components_250K} compares the role phonon-assisted scattering and coherent/incoherent Dexter coupling processes on the early CD dynamics. 
Here, at 250~K, phonon-assisted scattering is also the main mechanism behind the initial fast decay of the A signal and the delayed build-up of the B signal, as shown in Supplementary Figure~\ref{fig:theory_components_250K}a. 
Coherent and incoherent Dexter coupling cannot explain the formation dynamics on the observed timescales (see  figure~\ref{fig:theory_components_250K}b).
While the initial fast decay of the A signal and the delayed build-up of the B signal is relatively insensitive to the temperature, the overall decay is not. 
In Supplementary Figure~\ref{fig:theory_components_250K}c we depict the individual contributions of the considered scattering mechanisms. 
Here, in contrary to the dynamics at 77~K shown in Figure~6c, main text, intervalley exchange alone displays already a substantial impact on the valley depolarization. 
The reason is an increased availability of momentum-direct \mbox{$K^{}_\Uparrow$-$K^{}_\uparrow$} (\mbox{$K^{\prime}_\Downarrow$-$K^{\prime}_\downarrow$}) occupations, where intervalley exchange can act upon, due to an increased thermal energy of the excitonic occupations.
However, similar to the 77~K-case, intervalley exchange alone is not sufficient, but an interplay with Dexter interaction 
is needed to obtain valley depolarization times close to the experimentally observed dynamics.


\bibliography{sn-bibliography}

\end{document}